\documentclass{article}
\usepackage[
    bottom=2.8cm,
    left=2cm,
    right=2cm
]{geometry}
\usepackage[utf8]{inputenc}
\usepackage[american]{babel}
\usepackage{standalone}
\usepackage[autostyle,autopunct=true]{csquotes}
\usepackage[
    labelfont={bf},
    labelsep=period,
    font={small,sf},
    singlelinecheck=off,
    tableposition=top
]{caption}
\usepackage[square,sort,comma,numbers]{natbib}
\usepackage{subcaption}
\usepackage{graphicx}
\usepackage{tcolorbox}
\usepackage{enumitem}
\usepackage[
    colorlinks=true,
    linkcolor=blue,
    citecolor=blue,
    urlcolor=blue,
    bookmarksopen=true,
    bookmarksopenlevel=1
]{hyperref}
\usepackage{bookmark}
\usepackage[title]{appendix}
\usepackage{etoolbox}
\usepackage{amsmath}
\usepackage{amssymb}
\usepackage{amsthm}
\usepackage{mathtools}
\usepackage{tikz}
\usepackage{titlesec}
\usepackage{placeins}
\usepackage{booktabs}
\usepackage[flushleft]{threeparttable}
\usepackage{algorithm}
\usepackage[noend]{algpseudocode}

\graphicspath{{./figures/}}

\newcommand{\mean}[1]{\ensuremath{\langle#1\rangle}}
\newcommand{\N}{\mathcal{N}}

\title{Structural measures of similarity and complementarity in complex networks}
\author{Szymon Talaga$^{1\ast}$ \and Andrzej Nowak$^{2,3}$}
\date{
    {\normalsize
    $^1{}$Robert Zajonc Institute for Social Studies, University of Warsaw,\\
    Stawki 5/7, 00-183 Warsaw, Poland. \\
    $^2{}$Faculty of Psychology, University of Warsaw,\\
    Stawki 5/7, 00-183 Warsaw, Poland.\\
    $^3{}$Department of Psychology, Florida Atlantic University,\\
    777 Glades Rd, Boca Raton, FL 33431, USA.
    \\[1em]
    $^\ast{}$Corresponding author; E-mail: \texttt{stalaga@uw.edu.pl}
    }
}

\titleformat{\section}%
    {\normalfont\Large\bfseries}{\thesection.}{1em}{}
\titleformat{\subsection}%
    {\normalfont\large\bfseries}{\thesubsection.}{1em}{}
\titleformat{\subsubsection}%
    {\normalfont\normalsize\bfseries}{\thesubsubsection.}{1em}{}

\begin{document}

\flushbottom
\maketitle

\begin{abstract}
\noindent
The principle of similarity, or homophily, is often used to explain
patterns observed in complex networks such as transitivity
and the abundance of triangles (3-cycles).
However, many phenomena from division of labor to protein-protein interactions
(PPI) are driven by complementarity (differences and synergy).
Here we show that the principle of complementarity is linked to the abundance
of quadrangles (4-cycles) and dense bipartite-like subgraphs.
We link both principles to their characteristic motifs and
introduce two families of coefficients of:
(1)~structural similarity, which generalize local clustering and closure
coefficients and capture the full spectrum of similarity-driven structures;
(2)~structural complementarity, defined analogously but based on quadrangles
instead of triangles.
Using multiple social and biological networks, we demonstrate that the
coefficients capture structural properties related to
meaningful domain-specific phenomena. We show that they allow distinguishing
between different kinds of social relations as well as measuring
an increasing structural diversity of PPI networks across the tree of life.
Our results indicate that some types of relations are better explained by complementarity 
than homophily, and may be useful for improving existing link prediction methods.
We also introduce a Python package implementing efficient algorithms for
calculating the proposed coefficients.
\end{abstract}

\thispagestyle{empty}

\section*{Introduction}

The structure of complex networks commonly reflects their
functional properties as well as mechanisms or processes that created them.
Seminal studies have shown that different systems, from neural networks to the
World Wide Web, tend to be characterized by the presence of statistically
over-represented small subgraphs, known as network
motifs~\cite{miloNetworkMotifsSimple2002,shen-orrNetworkMotifsTranscriptional2002,tranCountingMotifsHuman2013}.
While one may expect different motifs to be related to particular
functions or properties of a given system, it is often not easy to determine
what they are exactly. In some cases and specific contexts, such as gene
regulatory networks, the roles played by different motifs may be revealed through
experimental studies~\cite{shen-orrNetworkMotifsTranscriptional2002,alonNetworkMotifsTheory2007}.
However, general principles that would explain the prevalence of specific
motifs across different application domains are still mostly unknown.

\begin{figure}[ht]
\centering
\includegraphics[width=.4\linewidth]{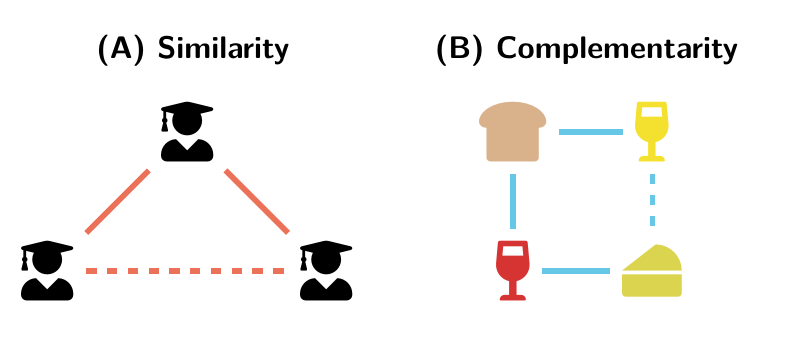}
\caption{%
Intuitive meaning of similarity and complementarity.
\textbf{(A)}~If three persons are similar (e.g.~they are all scientists)
and we know that one of them (top) knows the other two (bottom) it is
quite likely that they know each other too (dashed line).
Thus, the relation is \textbf{transitive}.
\textbf{(B)}~If one wine (red) goes well with (is complementary to) bread
and cheese and another wine (white) also goes well with the bread,
then it is likely that it is a good match for the cheese too (dashed line).
However, this does not imply that both wines will be drunk together,
so the relation is \textbf{not transitive}.
}
\label{fig:intuition}
\end{figure}

An important exception is the widely-known abundance of triangles (3-cycles)
in many types of real-world networks, which has been shown to be a structural
signature of transitive relations driven by similarity between nodes in some
(possibly latent) metric space~\cite{bogunaNetworkGeometry2021,bogunaSmallWorldsClustering2020,krioukovClusteringImpliesGeometry2016}.
The importance of similarity and its impact on the structure of social networks
has been recognized in sociology for a long time, as it is linked to homophily
and triadic closure~\cite{marsdenHomogeneityConfidingRelations1988,mcphersonBirdsFeatherHomophily2001,kossinetsOriginsHomophilyEvolving2009,asikainenCumulativeEffectsTriadic2020,talagaHomophilyProcessGenerating2020}.
While it is usually hard to disentangle their
effects~\cite{anagnostopoulosInfluenceCorrelationSocial2008,aralDistinguishingInfluencebasedContagion2009},
these two processes are also inherently linked as they lead to
high structural equivalence~\cite{newmanNetworksIntroduction2010} 
between connected nodes. In other words, in similarity-driven
systems two adjacent nodes are likely to share a lot of neighbors
(Fig.~\ref{fig:intuition}A), and this implies the abundance of triangles
and a latent geometric
structure~\cite{papadopoulosNetworkGeometryInference2015,krioukovClusteringImpliesGeometry2016}.

An alike, even if less known, phenomenon is the connection between the abundance of 
quadrangles (4-cycles) and networks with so-called functional 
structure~\cite{mattssonFunctionalStructureProduction2021},
in which two nodes interact not because they are similar, but rather because
one of them is similar (in some salient way) to the neighbors of the
other~\cite{kovacsNetworkbasedPredictionProtein2019}. This linkage principle leads
to markedly different local connectivity structures than those found in networks
dominated by triangles (e.g.~typical social networks) and is characteristic for relations
driven by complementarity, or differences and synergies, between the features of connected
elements~\cite{kovacsNetworkbasedPredictionProtein2019,mattssonFunctionalStructureProduction2021,jiaMeasuringQuadrangleFormation2021}.

This observation is important as many phenomena across different application domains,
from cooperation, business interactions and division of labor~\cite{gulatiSocialStructureAlliance1995,chungComplementarityStatusSimilarity2000,riveraDynamicsDyadsSocial2010,xieSkillComplementarityEnhances2016,dopferUpwardDownwardComplementarity2016,mattssonFunctionalStructureProduction2021}
to the quality of romantic relationships~\cite{markeyRomanticIdealsRomantic2007},
consumer choices~\cite{tianExtractingComplementsSubstitutes2021}
and at least some types of protein-protein binding~\cite{kovacsNetworkbasedPredictionProtein2019},
may indeed be better explained by the principle of complementarity than similarity.
For instance, two types of wine may be often bought together with the same
kinds of bread and cheese, but rarely both of them will occur in the same
transaction. In other words, in this situation a wine is complementary to the
bread and cheese, but not to the other wine (Fig.~\ref{fig:intuition}B).
More generally, complementarity can be seen as a particular interpretation of
the principle of heterophily, which is a preference for connecting to others
who are different with respect to some salient
attributes~\cite{riveraDynamicsDyadsSocial2010}.

Here we show that the principle of complementarity, unlike the more general notion of heterophily,
has a straightforward geometric interpretation which links it to quadrangles as its characteristic
motif, in the same way as the intrinsic geometry of similarity links it to triangles.
We also show that under a particular quadrangle definition (4-cycle without diagonal shortcuts)
the principle of complementarity is connected to locally dense subgraphs of
high bipartivity~\cite{holmeNetworkBipartivity2003}, which, again, is analogous to how the 
abundance of triangles implies the presence of dense unipartite subgraphs.
More generally, we argue that both similarity and complementarity are important relational
principles shaping the structure of networks across different application domains
and provide a generic explanation for some of the prevalent structural patterns observed
in many real-world systems. 

In order to formalize our analysis, we first define a general family
of similarity coefficients measuring the abundance of triangles at the levels
of individual nodes and edges as well as entire graphs. The coefficients
generalize the notions of local clustering and
closure~\cite{wattsCollectiveDynamicsSmallworld1998,yinLocalClosureCoefficient2019}
and therefore capture the full spectrum of transitive, similarity-driven
structures. Then, starting from a simple geometric model of complementarity we
follow the same logic as in the case of similarity and define an analogous
family of complementarity coefficients measuring the abundance of quadrangles.

We will call the proposed measures \textit{structural coefficients}
because they will not be defined with respect to node attributes, latent or
observed, but to how different nodes are embedded in the network. 
Moreover, they will not measure (dis)similarity between nodes,
as this problem is usually addressed by measures of 
\textit{structural equivalence}~\cite{newmanNetworksIntroduction2010}. 
Instead, structural coefficients will measure the extent
to which any given edge, node or graph is compatible with the principle
of similarity or complementarity. However, to facilitate the interpretation
we will also show how the proposed notions of structural similarity and 
complementarity are related to structural equivalence.

We study the behavior of structural coefficients in some of the most
important random graph models as well as multiple real-world social and
biological networks. We demonstrate that they are related to meaningful
domain-specific phenomena and can be used to distinguish between different
types of networks. In particular, 
using a collection of comparable real-world networks measuring friendship and health advice ties,
we show that structural coefficients discriminate effectively between social relations 
driven by similarity and complementarity, which provides evidence for the theoretical
validity of our approach.
We also demonstrate how the coefficients may be used to measure the increasing structural
diversity of protein-protein interactions (PPI) across the tree of life based on hundreds 
of interactome networks of different organisms.

Our work complements the rich literature on network motifs, network geometry
and local connectivity structures
as well as introduces principled theory and methods linking different
types of relations to their observable structural signatures. 
We argue that the customary assumption of homophily is not adequate for some types
of social relations, which are better explained by complementarity,
and provide tools for identifying such systems,
bringing more nuance to the field of social network analysis. Moreover,
the framework we propose could be, in principle, used for improving existing
link prediction methods by helping to determine when the assumption of 2-path
(L2/triadic) or 3-path (L3/tetradic) closure~\cite{kovacsNetworkbasedPredictionProtein2019}
is more appropriate.
Last but not least, all methods introduced in this paper are implemented in a
Python package called \texttt{pathcensus} (see Materials and Methods).

\subsection*{Notation \& technical remarks}

In this paper we consider simple undirected and unweighted graphs $G = (V, E)$
with no self-loops.
We use $n = |V|$ and $m = |E|$ to denote the numbers of nodes
and edges in $G$ respectively. Elements of the adjacency matrix of $G$
will be denoted by $a_{ij}$ and assumed to be equal to $1$ if the edge
$(i, j)$ exists and $0$ otherwise. For any node $i \in V$ we denote its degree
by $d_i$ and its $k$-hop neighborhood by $\N_k(i)$, in particular $1$-hop
neighborhood will be denoted by $\N_1(i)$
(a $k$-hop neighborhood consists of nodes connected to $i$ by a shortest path of length $k$).
Moreover, we will use $n_{ij} = |\N_1(i) \cap \N_1(j)|$ to denote
the number of shared neighbors between nodes $i$ and $j$.
Averaged quantities will be denoted by
diamond brackets. For instance, $\mean{d_i}$ will denote average node degree.

\subsection*{Structural equivalence}

We briefly introduce the notion of \textit{structural equivalence}, to which we will refer
at multiple points throughout the paper. Structural equivalence is a measure of the extent 
to which two nodes are similarly embedded in a network. 
It can be defined in multiple ways, but all definitions try to quantify 
similarity between $1$-hop neighborhoods of two 
nodes~\cite{newmanNetworksIntroduction2010}. 
Here we will follow a common approach and define structural equivalence in terms of 
Sørenson Index or normalized Hamming similarity:
\begin{equation}\label{eq:hamming}
    H_{ij} = \frac{2n_{ij}}{d_i + d_j}
\end{equation}
which is also often used as an index for predicting missing links 
(under the assumption of triadic closure)~\cite{srilathaSimilarityIndexBased2016}.
Crucially, the notion of structural equivalence applies to pairs of nodes 
(not necessarily connected) and is concerned with the degree of (dis)similarity 
of their $1$-hop neighborhoods. This is in contrast to structural coefficients we propose,
which are descriptors of edges, nodes or graphs capturing the degree to which they are
compatible with the logic of similarity or complementarity.

\section*{Theory and Definitions}

Here we present the proposed theory of structural similarity and complementarity and
introduce all the main definitions that will be used throughout the paper.
We first discuss structural similarity and its nodewise and global coefficients and then
define the analogous complementarity coefficients. In the second part of the section
we introduce edgewise measures and discuss use them to discuss the connection between
similarity, complementarity and structural equivalence.

\subsection*{Structural similarity}

\begin{figure}[t]
\centering
\includegraphics[width=\textwidth]{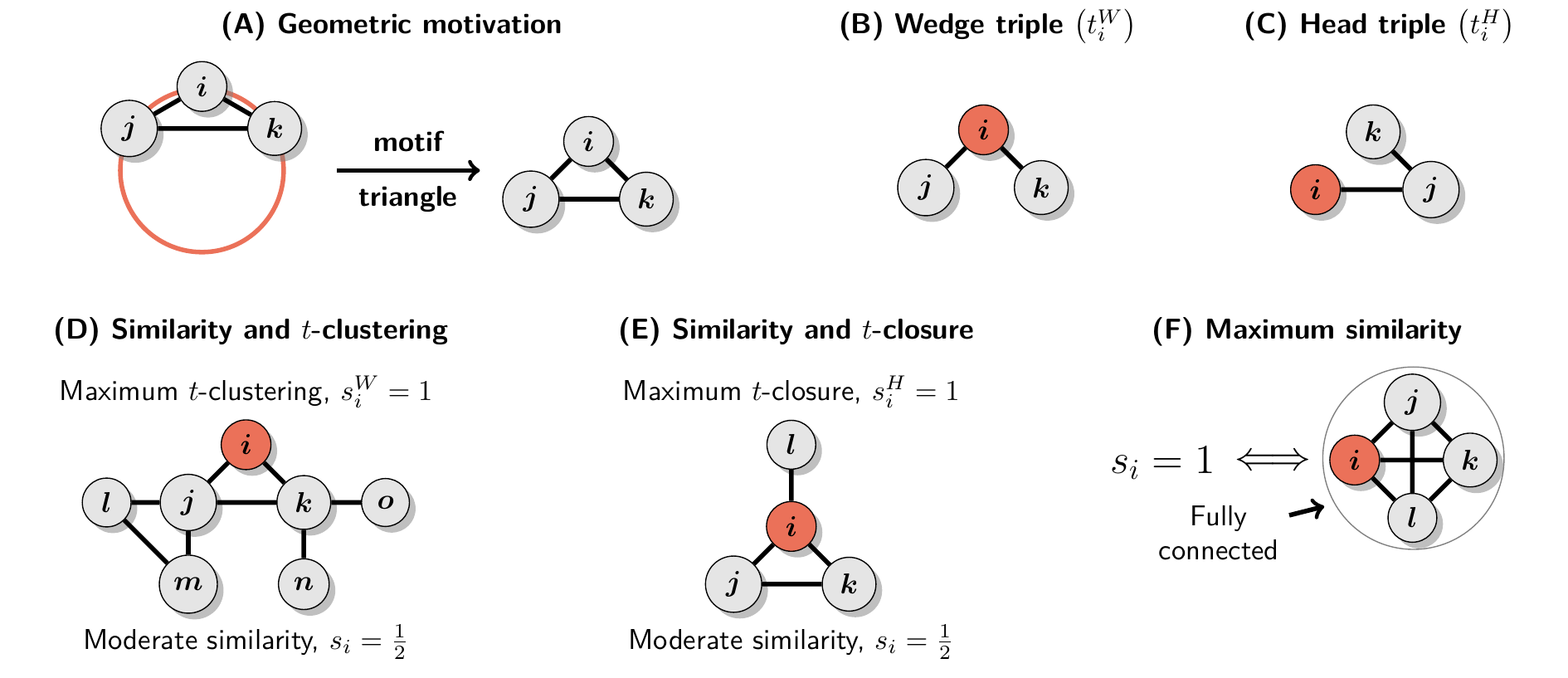}
\caption{%
Geometric motivation and the main properties of structural
similarity coefficients.
\textbf{(A)}~Metric structure induced by similarity implies
transitivity of relations and the abundance of triangles.
\textbf{(B~and~C)}~Wedge and head triples.
\textbf{(D)}~Local clustering can be maximized even when neighbors of the
focal node are very differently embedded within the network, while $s_i$
is sensitive to this kind of non-transitivity.
\textbf{(E)}~Local closure can be maximized even for nodes with sparse
1-hop neighborhoods if they are star-like as neighbors with degree one
do not generate any head triples. On the other hand, $s_i$ is sensitive
to this violation of transitivity.
\textbf{(F)}~Necessary and sufficient conditions for maximum
structural similarity.
}
\label{fig:sim}
\end{figure}

It is common to think about similarity in terms of distance between
different objects in a feature space. Hence, the motivating geometric
model for similarity-driven relations posits that nodes are positioned in some
metric space and the probability of observing a link between them is a
decreasing function of the corresponding distance. Such a generic model can be
seen as an instance of the class of Random Geometric Graphs
(RGG)~\cite{bogunaSmallWorldsClustering2020,talagaHomophilyProcessGenerating2020}.
The crux is that this very general formulation is enough to guarantee the
abundance of triangles (3-cycles) (see Fig.~\ref{fig:sim}A).

Thus, a good starting point for our endeavor is
\textit{local clustering coefficient}~\cite{wattsCollectiveDynamicsSmallworld1998},
of which value for a node $i$ will be denoted by $s^W_i$.
It is a classical network measure of the density of the $1$-hop neighborhood
(ego-network) of $i$ and is defined as:
\begin{equation}\label{eq:tclust}
    s^W_i 
    = \frac{2T_i}{t^W_i}
    = \frac{\sum_{j,k}a_{ij}a_{ik}a_{jk}}{d_i(d_i-1)}
\end{equation}
where $T_i$ is the number of triangles including $i$ and
$t^W_i$ is the number of wedge triples centered at $i$ or
$2$-paths with $i$ in the middle (Fig.~\ref{fig:sim}B).
Crucially, $s^W_i \in [0, 1]$ and is equal
to $1$ if and only if $\N_1(i)$ forms a fully connected network.
In sociological terms, it measures the extent to which
\textit{my friends are friends with each other}. Note, however, that this is
only one side of the triadic closure process as it corresponds to
the closing of the loop between friends of the focal node~$i$.
The other part is about the loop between $i$ and friends of its friends and
local clustering coefficient does not capture it.

To address this issue an alternative
\textit{local closure coefficient}~\cite{yinLocalClosureCoefficient2019}
has been proposed more recently:
\begin{equation}\label{eq:tclosure}
    s^H_i 
    = \frac{2T_i}{t^H_i}
    = \frac{\sum_{j,k}a_{ij}a_{ik}a_{jk}}{\sum_j a_{ij}(d_j-1)}
\end{equation}
where $t^H_i$ is the number of head triples originating from $i$, that is,
$2$-paths starting at $i$ (Fig.~\ref{fig:sim}C).
It is also in the range of $[0, 1]$ and attains the
maximum value if and only if no neighbor of $i$ is adjacent to a node which
is not already in $\N_1(i)$. In other words, when $s^H_i = 1$ a random walker
starting at $i$ will never leave $\N_1(i)$. Thus, local closure coefficient
measures the extent to which \textit{friends of my friends are my friends},
that is, it is a measure of triadic closure between the focal node $i$
and neighbors of its neighbors. As a result, it captures exactly that what
local clustering is blind to. Since the local clustering and closure
coefficients are based on triples we will later refer to them as
$t$-clustering and $t$-closure respectively.

The two coefficients complement each other, so it is justified to combine
them in a single measure. We now propose such a measure which we will call
\textit{structural similarity coefficient}:
\begin{equation}\label{eq:simcoef}
    s_i
    = \frac{4T_i}{t^W_i + t^H_i}
    = \frac{t^W_is^W_i + t^H_is^H_i}{t^W_i + t^H_i}
\end{equation}
Note that $s_i$ is equal to the fraction of both wedge and head triples
including $i$ which can be closed to make a triangle. It is also equivalent to
a weighted average of $s^W_i$ and $s^H_i$, which implies that
$\min(s^W_i, s^H_i) \leq s_i \leq \max(s^W_i, s^H_i)$.
As we show later, this makes $s_i$ a more general descriptor
of local structure than $s^W_i$ or $s^H_i$ alone (cf.~Section: Configuration model).
Moreover, since $s^W_i = 1$ if and only if $\N_1(i)$ is fully connected
and $s^H_i = 1$ if there are no links leaving $\N_1(i)$ then it must be that
$s_i = 1$ if and only if $i$ belongs to a fully connected network.
Fig.~\ref{fig:sim} provides a summary of the motivation and main properties of
$s_i$, including examples of when $t$-clustering and $t$-closure coefficients
are maximal while structural similarity is only moderate
(Figs. \ref{fig:sim}D and \ref{fig:sim}E).
Crucially, unlike local clustering and closure, structural similarity
is a comprehensive measure of the density of triangles around a node $i$ and
therefore captures the full spectrum of local structures implied by the
transitivity of similarity-driven relations. 
Moreover, it is defined for all nodes contained within components with at least 3 nodes.
This is in contrast to local clustering which is not defined for nodes with $d_i = 1$.

\subsubsection*{Global similarity}

From the global perspective both local clustering and local closure lead
to the same conclusion that the corresponding global measure is just the
fraction of triples that can be closed to make a
triangle~\cite{yinLocalClosureCoefficient2019}.
This implies that the same quantity is also the proper global measure of
the extent to which relations are driven by similarity. In other words,
\textit{global similarity coefficient} is equal to the standard global
clustering coefficient and can be defined as:
\begin{equation}\label{eq:sim-global}
    s = \frac{3T}{\sum_i d_i(d_i-1)}
\end{equation}
where $T$ is the total number of triangles and the denominator counts the
number of triples.

Note that it is indeed a reasonable measure of similarity-driven relations 
as it is maximized only when a network is fully connected, 
so all nodes are structurally redundant and each can be removed
without affecting the overall connectivity.

\subsection*{Structural complementarity}

First, let us consider an intuitive meaning of complementarity. We posit that
two objects are complementary when their features are different but in a
well-defined synergistic way. As we will see, this additional synergy
constraint is crucial. However, before we discuss this further let us note
that in the case of similarity an analogous constraint is built-in by design.
For any point there is always only one point minimizing the
distance (maximizing similarity) and it is the point itself. 
In other words, any object is most similar to itself.
As a result, there is a well-defined notion of maximal similarity.

On the other hand, the case of difference is more complicated.
To make our argument more concrete, let the feature space be
$\mathbb{R}^k$ with $k \geq 1$. Now, it is easy to see that
for any two points $p$ and $r$ at a distance $d(p, r)$ we can find a third
point $s$ such that $d(p, s) > d(p, r)$. In other words, for any point $p$
there is no well-defined point at the maximum distance. Thus, complementarity
cannot be defined in terms of arbitrary differences. Intuitively, defining it
in terms of a simple unconstrained heterophily inevitably leads to the
conclusion that for any object there is an infinite variety of more and more
complementary (different) objects, which clearly does not map well on the
common understanding of the notion of complementarity. Thus, we need a
definition with the same property as in the case of similarity, that is,
one yielding a sequence of ever smaller sets of more and more complementary
elements converging to a single well-defined point in the limit of maximum
complementarity.

Note that the above abstract argument can be related to known
complementarity-driven systems in a rather straightforward manner.
For instance, a key and a lock are complementary not because
they are just different in an arbitrary fashion, but because they differ in a
very specific way by being structural negatives of each other. Similarly,
division of labor in modern societies is based on complex synergies between
capabilities of different individuals and organizations.

Thus, we argue that complementarity should be defined in terms of distance
maximization but with additional constraints ensuring that for any point in the
feature space there is only one point at the maximum distance. This can be
achieved in several ways, but to keep things simple we will
focus on one particularly straightforward solution.

\begin{figure}[t]
\centering
\includegraphics[width=\textwidth]{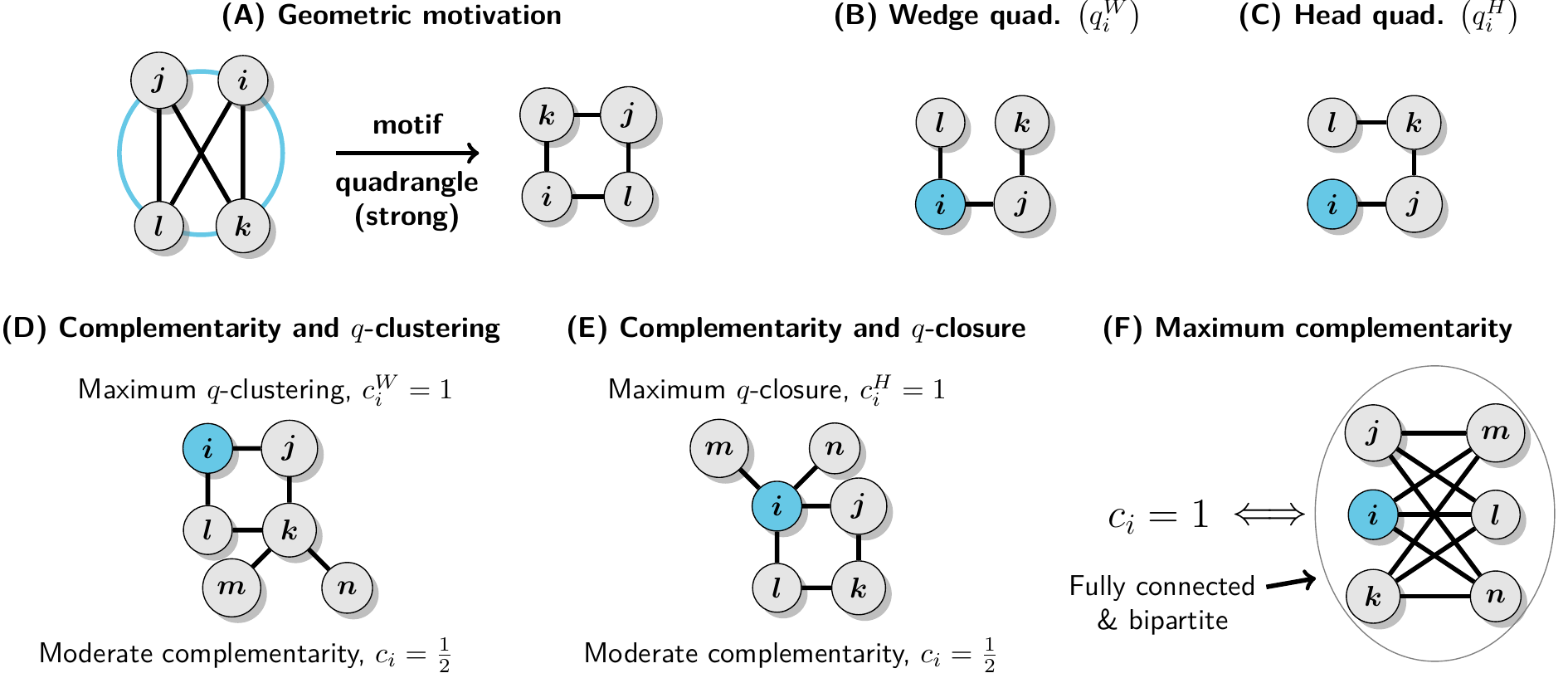}
\caption{%
Geometric motivation and the main properties of structural
complementarity coefficients.
\textbf{(A)}~On~the surface of a (hyper)sphere for
each point there is only a single other point at the maximum distance,
so complementarity based on distance maximization must lead to
the abundance of strong (chordless) quadrangles and locally dense
bipartite-like subgraphs. 
\textbf{(B~and~C)} Wedge and head quadruples.
\textbf{(D)}~Local $q$-clustering can be
maximized even when some 2-hop neighbors (node $k$ on the figure)
of the focal node connect to nodes which are not in $\N_1(i)$,
while the $c_i$ is sensitive to this deviation from the principle of
complementarity.
\textbf{(E)}~Local $q$-closure can be maximized even for nodes with sparse
1-hop neighborhoods if they are star-like as neighbors with degree one
do not generate any head quadruples.
\textbf{(F)}~Necessary and sufficient conditions for maximum
structural complementarity.
}
\label{fig:comp}
\end{figure}

We consider nodes as placed on the surface of a $k$-dimensional (hyper)sphere
with $k \geq 1$. In this setting for each point there is only a single point
at the maximum distance and the maximum distance is the same for all points.
Now, if nodes connect preferentially to others who are far away, we obtain a
model analogous to similarity, but the connections of a node are not
concentrated in its vicinity but instead on the other side of the space.
From this it follows that any two connected nodes $i$ and $j$ will not share a
lot of neighbors, so triangles will be rare, but instead the $1$-hop
neighborhood of $i$ should be approximately equal to the $2$-hop neighborhood
of $j$ and \textit{vice versa}, that is,
$\N_1(i) \approx \N_2(j)$ and $\N_2(i) \approx \N_1(j)$.
Such a spatial structure inevitably leads to the abundance of
quadrangles (4-cycles) and the presence locally dense bipartite-like
subgraphs~(Fig.~\ref{fig:comp}A).
There are, of course, alternative and more general ways in which geometric models of 
complementarity-driven relations can be defined 
(see Ref.~\cite{kitsakLatentGeometryComplementarityDriven2020} for an excellent example),
but distance maximization on a sphere provides a good minimal model highlighting 
the connection between complementarity, bipartivity and quadrangles.

Depending on the context different authors may refer to slightly different
objects when using the term \textit{quadrangle}. Namely, a quadrangle
may contain up to two chords or diagonal links.
Here we will consider only quadrangles without any chords, which we will call
\textit{strong quadrangles}. This choice follows, of course, from the proposed
geometric model and the fact that only strong quadrangles are characteristic
for dense bipartite-like graphs, which should not have many odd cycles.

Now we can start defining coefficients measuring relations driven by
complementarity. As previously, we begin with a local clustering coefficient,
which will be called $q$-clustering. It is defined analogously,
but this time in terms of quadrangles and wedge quadruples, that is,
$3$-paths with the focal node $i$
at the second position (Fig.~\ref{fig:comp}B):
\begin{equation}\label{eq:qclust}
    c^W_i 
    = \frac{2Q_i}{q^W_i}
    = \frac{\sum_{j \neq i}a_{ij}\sum_{k \neq i, j}a_{ik}(1-a_{jk})\sum_{l \neq i,j,k}a_{kl}a_{jl}(1-a_{il})}
    {\sum_j a_{ij}[(d_i-1)(d_j-1)-n_{ij}]}
\end{equation}
where $Q_i$ is the number of strong quadrangles incident to the focal node $i$
and $q^W_i$ is the number of wedge quadruples it belongs to. Note that we
consider only quadruples with $i$ at the second position, such as
$(l, i, j, k)$ but not $(k, j, i, l)$, in order to avoid
double counting and make the number of wedge and head quadruples per quadrangle
equal. Intuitively, it quantifies the extent to which the local environment
of $i$ is bipartite-like and its neighbors are structurally equivalent
to each other.

Local $q$-closure coefficient is defined in the same way as the fraction
of head quadruples originating from $i$ (Fig.~\ref{fig:comp}C)
that can be closed to make a (strong) quadrangle:
\begin{equation}\label{eq:qclosure}
    c^H_i 
    = \frac{2Q_i}{q^H_i}
    = \frac{\sum_{j \neq i}a_{ij}\sum_{k \neq i, j}a_{ik}(1-a_{jk})\sum_{l \neq i,j,k}a_{kl}a_{jl}(1-a_{il})}
    {\sum_{j \neq i}a_{ij}\sum_{k \neq i, j}a_{jk}(d_k - 1 - a_{ik})}
\end{equation}
where $q^H_i$ is the number of head quadruples starting at $i$.
Conceptually, it measures the extent to which the local environment of $i$
is bipartite-like and $i$ is structurally equivalent to its $2$-hop neighbors.

We can now define \textit{structural complementarity coefficient}
as the fraction of quadruples including the focal node $i$ which can be closed
to make a (strong) quadrangle which is equivalent to a weighted
average of $q$-clustering and $q$-closure:
\begin{equation}\label{eq:comp}
    c_i
    = \frac{4Q_i}{q^W_i + q^H_i}
    = \frac{q^W_ic^W_i + q^H_ic^H_i}{q^W_i + q^H_i}
\end{equation}
Note that again we have that
$\min(c^W_i, c^H_i) \leq c_i \leq \max(c^W_i, c^H_i)$,
so $c_i$ is always bounded between its constitutive clustering and closure
coefficients. 
This implies that $c_i$ is a more general descriptor than $c^W_i$ or $c^H_i$ alone
(cf.~Section: Configuration model).
Moreover, the interpretations of $q$-clustering and $q$-closure
jointly imply that $c_i = 1$ if and only if the focal node $i$ belongs to a
fully connected bipartite network. Fig.~\ref{fig:comp} presents a summary of
the most important terms and facts related to $c_i$.

The geometric model underlying the definition
of $c_i$ indeed justifies the interpretation in terms of complementarity
or synergy. Nodes are more likely to be connected when they are far away
in the feature space, meaning that they have different properties
which can be possibly combined in a synergistic manner. Crucially,
the mesoscopic network structure that is implied by this model is also related
to complementarity in a straightforward manner. Bipartite networks are
representations of complementarity-driven systems \textit{par excellence}
as they consist of two types of nodes and allow only for connections between
them. 
Thus, $c_i$, being a measure of local bipartivity, is indicative of the degree to which
the local environment of a node resembles such a complementarity-driven system.

However, our measure of structural complementarity, while closely related to measures
of network  bipartivity~\cite{holmeNetworkBipartivity2003,estradaSpectralMeasuresBipartivity2005},
is also different in at least two important respects. Firstly, unlike bipartivity
measures, structural complementarity captures both local bipartivity and density.
This is important because even a high degree of bipartivity alone is not a signature
of complementarity, since random tree-like structures are also relatively bipartite-like
(as evident in Fig.~3a in Ref.~\cite{holmeNetworkBipartivity2003} where bipartivity
coefficients, $b_1$ and $b_2$, are much higher than the minimal value of $1/2$
even for networks with very low values of $r_1$ parameter which are effectively 
Erdős–Rényi random graphs). 
Secondly, bipartivity measures are typically 
global~\cite{holmeNetworkBipartivity2003,estradaSpectralMeasuresBipartivity2005},
while structural complementarity coefficients can be defined for edges, nodes and entire 
graphs (we note, however, that spectral bipartivity can be defined also for individual
nodes~\cite{estradaProteinBipartivityEssentiality2006}).

Furthermore, structural complementarity coefficient follows closely
the definitions of \textit{i-quad} and \textit{o-quad} coefficients proposed in 
Ref.~\cite{jiaMeasuringQuadrangleFormation2021}. However, it also differs in two
important respects. Firstly, it combines both the perspective of wedge (\textit{i-quad})
and head (\textit{o-quad}) quadruples. 
As we show later (Section: Configuration model), this makes $c_i$ a more general
descriptor of local structure and the density of quadrangles, even if 
for some specific research questions clustering or closure (\textit{i-quad} or \textit{o-quad}) coefficients may still be more appropriate.
Secondly, it is based on the notion of strong (chordless) quadrangles instead of the weaker 
notion allowing for any number of chordal edges. This is necessary for ensuring the direct 
connection to bipartivity. However, it comes at a cost of making structural complementarity 
coefficient more sensitive to noise 
(as strong quadrangles can be easily destroyed by a single erroneous chordal edge)
and less capable of detecting structures deviating from the strict assumption of local
bipartivity. Of course, $c_i$ can be redefined using weak quadrangles, which would lead
to a measure equivalent to a weighted average of \textit{i-quad} and \textit{o-quad} coefficients.
However, developing a proper interpretation of weak quadrangles
\textit{vis-à-vis} the principles of similarity and complementarity would require a non-negligible
amount of additional theoretical and mathematical work, which is outside the scope
of this paper. Nonetheless, weak quadrangles may have some interesting applications as, for instance,
they seem to be connected to the theory of large quasirandom graphs, of which structure is determined
by the amount of general 4-cycles~\cite{lovaszLargeNetworksGraph2012}.
Thus, we plan to address this problem in the future.

When applied to bipartite networks the quadrangle-based
measures can be seen as a generalization of the bipartite clustering
coefficient(s)~\cite{zhangClusteringCoefficientCommunity2008,opsahlTriadicClosureTwomode2013a}.
However, the crux is that our structural complementarity
coefficients can be applied to unipartite networks in order to quantify
jointly local bipartivity and density, which together are indicative of
complementarity-driven relations.

\subsubsection*{Global complementarity coefficient}

From the global perspective of an entire network there is of course no
difference between wedge and head quadruples. Hence, the global coefficient
can be defined simply as:
\begin{equation}\label{eq:comp-global}
    c = \frac{4Q}{\sum_{i,j} a_{ij}[(d_i - 1)(d_j - 1) - n_{ij}]}
\end{equation}
where $(i, j) \in E$ and $Q$ is the total number of quadrangles with no chords.
The denominator counts the total number of quadruples. Note that $c = 1$
if and only if the graph as such is fully connected and
bipartite. This agrees with the intuition as this is exactly the structure
one should expect in a system composed of two classes of elements in which
each element in one class is perfectly complementary to every element in the
other.

\subsection*{Edgewise measures and structural equivalence}

\subsubsection*{Similarity}

\begin{figure}[th]
\centering
\includegraphics[width=.9\textwidth]{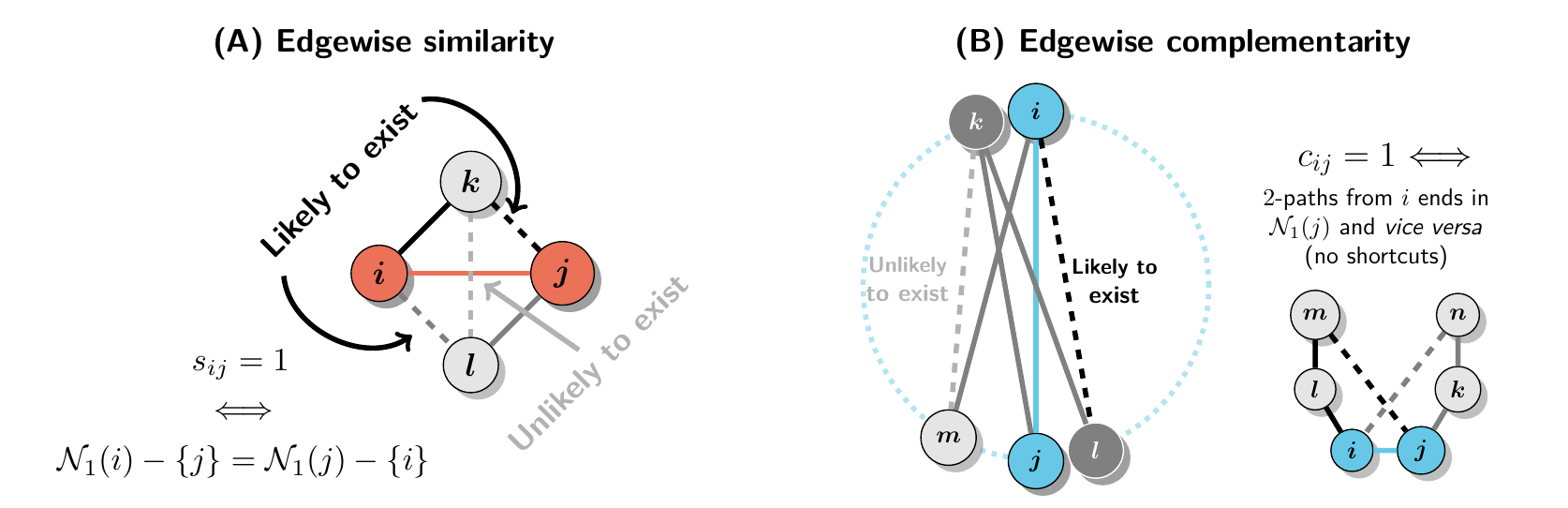}
\caption{%
Interpretation of the edgewise structural coefficients.
Focal edges are marked with colors (red or blue)
and paths starting from the focal $(i, j)$ edge are black and dark grey.
Solid lines denote paths (triples or quadruples) and dashed lines correspond
to closing edges that form triangles or quadrangles.
\textbf{(A)} Logic of edgewise similarity and the necessary and sufficient conditions
for maximum $s_{ij}$. If $(i, j)$ edge is driven by similarity then any neighbor 
of either $i$ or $j$ ($k$ and $l$ on the figure) should be near $i$ (or $j$) in the latent
space making it likely that it links to the other member of the $(i, j)$
pair too. On the other hand, $k$ and $l$ may still be quite far away
so the link between them is unlikely to exist.
\textbf{(B)} Logic of edgewise complementarity and necessary and sufficient conditions
for maximum $c_{ij}$. If $(i, j)$ edge is driven by complementarity then any $2$-hop
neighbor of $j$ ($i$), such as $l$ ($k$) on the figure, should be a $1$-hop
neighbor of $i$ ($j$). The quadruple $(i, j, k, l)$ corresponds to such a situation. 
On the other hand, any pair of neighbors of $i$ and $j$ correspondingly may be located 
in the latent space close enough to each other as to make a tie between them unlikely
(as it happens for nodes $m$ and $k$ on the figure).
}
\label{fig:edgewise}
\end{figure}

Edgewise structural similarity coefficient is equal to the ratio of triangles 
including nodes $i$ and $j$ and the total number of $2$-paths traversing the $(i, j)$ edge
(Figs.~\ref{fig:sim}G and~\ref{fig:edgewise}A). In other words, it is equivalent to the number
of shared neighbors relative to the total number of neighbors of $i$ and $j$,
excluding $i$ and $j$ themselves:
\begin{equation}\label{eq:sim-edges}
    s_{ij}
    = \frac{2T_{ij}}{t^W_{ij} + t^H_{ij}}
    = \frac{2n_{ij}}{d_i + d_j - 2}
\end{equation}
where $T_{ij}$ is the number of triangles including
$i$ and $j$, $t^W_{ij}$ is the number of $(k, i, j)$ and $t^H_{ij}$
of $(i, j, k)$ triples. Importantly, $s_{ij}$ is symmetric since
$T_{ij} = T_{ji}$ and $t^W_{ij} = t^H_{ji}$.

Note that $s_{ij}$ is closely related to 
Hamming similarity defined in Eq.~(\ref{eq:hamming})
and differs only in the $-2$ term in the denominator which accounts
for the fact that $i$ and  $j$ are known to be connected. 
Together with the fact that nodewise coefficient $s_i$ is a weighted average
of the corresponding edgewise coefficients, or $\min_j{s_{ij}} \leq s_i \leq \max_j{s_{ij}}$
for $j \in \N_1(i)$, this implies that $s_i$
can be seen as a proxy for the extent to which $i$ is structurally equivalent to its own
neighbors. 

More concretely, it can be shown that:
\begin{equation}\label{eq:sim-bounds}
    \min_j H_{ij}
    < s_i \leq
    \max_j\left(H_{ij}\frac{d_i + d_j}{d_i + d_j - 2}\right)
\end{equation}

In other words, high (low) $s_i$ implies the existence of highly (lowly) structurally equivalent neighbor(s). Crucially, this also explains why structural similarity is inherently
linked to transitivity. If neighbors of $i$ are highly structurally equivalent to it,
then it must be likely that if $i \sim j$ and $j \sim k$ then $i \sim k$
or if $i \sim j$ and $i \sim k$ then $j \sim k$.
The proof of the above statements is presented in the Supplementary Information 
(SI: Similarity and structural equivalence).

\subsubsection*{Complementarity}

Edgewise structural complementarity coefficient is defined as:
\begin{equation}\label{eq:comp-edges}
    c_{ij} = \frac{2Q_{ij}}{q^W_{ij} + q^H_{ij}}
\end{equation}
where $Q_{ij}$ is the number of quadrangles including nodes $i$
and $j$, $q^W_{ij}$ is the number of $(j, i, k, l)$ and $q^H_{ij}$ of
$(i, j, k, l)$ quadruples. Again, $Q_{ij} = Q_{ji}$ and $q^W_{ij} = q^H_{ji}$
so $c_{ij}$ is symmetric.

This way $c_{ij}$ can be seen as a joint measure of
bipartivity around an $(i, j)$ edge and structural equivalence between $i$
and $1$-hop neighbors of $j$ and \textit{vice versa}. It measures the extent
to which $\N_2(i) \approx\ \N_1(j)$ and $\N_1(i) \approx \N_2(j)$ without requiring
dense connections between the $1$-hop and $2$-hop neighborhoods of $i$ and $j$.
This is in analogy to edgewise similarity which measures only the extent to which
$\N_1(i) \approx \N_1(j)$ without considering the density of connections between
the neighbors of $i$ and $j$ as this would be a higher-order property
unrelated to whether an edge is driven by similarity or not
(see Fig.~\ref{fig:edgewise} for details).

The connection to structural equivalence is slightly more complicated
in the case of complementarity and necessitates an introduction of an additional quantity.
For a connected triple $(k, i, j)$ we define \textit{Asymmetric Excess Sørenson Index}:
\begin{equation}\label{app:eq:excess-sorenson}
    H_{kj|i} = \frac{n_{jk}-1}{d_k - 1 - a_{jk}}
\end{equation}
which measures how many of the connections of $k$ are also shared by $j$
while disregarding edges $(i, k)$, $(i, j)$ and $(j, k)$.
Note that the excess degree  of $k$ is used in the denominator as the $(i, k)$
link needs to be ignored. Moreover, $a_{jk}$ term accounts for the possible
presence of the $(j, k)$ link. Finally, $1$ is subtracted from $n_{jk}$ to
account for the fact that $i$ is a shared neighbor of $j$ and $k$.

Now, using the fact that $c_i$ is a weighted average of $c_{ij}$'s,
or $\min_j c_{ij} \leq c_i \leq \max_j c_{ij}$, it can be shown that:
\begin{equation}\label{eq:comp-bounds}
    0 \leq c_i \leq
    \max_{j, k, l}\left(H_{kj|i}, H_{li|j}\right)
\end{equation}
where $j \in \N_1(i)$, $k \in \N_1(i)-\{j\}$ and $l \in \N_1(j)-\{i\}$
(see the proof in SI: Complementarity and structural equivalence).

In other words, $c_i$ is bounded from above by the maximum
Asymmetric Excess Sørenson Index between any two of its neighbors or itself
and any neighbor of its neighbors.
Intuitively, high complementarity can exist only in the presence
of high structural equivalence between neighbors of $i$ as well as $i$ and
neighbors of its neighbors.

Crucially, this explains in what sense complementarity-driven relations
are not transitive but yet localized. The principle of complementarity
enforces both the lack of connections between 1-hop neighbors of $i$ as well as
a degree of structural equivalence between them. This in turn induces a particular
kind of correlations between the connections of $i$ and its 1- and 2-hop neighbors
which at the same time do not imply transitivity of relations.

\section*{Results}

Here we present the results of four case studies analyzing the behavior of structural
coefficients in random graph models and using them to answer specific research
questions based on several empirical datasets.

\subsection*{Structural coefficients in random graphs}

\subsubsection*{Erdős–Rényi model}

In the Erdős–Rényi (ER)
model~\cite{erdosRandomGraphs1959} the expected global similarity,
which is of course equivalent to global clustering,
is simply $\mathbb{E}[s] = p$, or equal to the probability that any edge exists.
This is a standard result that follows from the fact that for any $(i, j, k)$
triple the closing $(i, k)$ edge always exists with
probability $p$~\cite{newmanNetworksIntroduction2010}.

We can use a similar argument to derive the expected value of
global complementarity coefficient in the ER model.
Let $(i, j, k, l)$ be any connected quadruple. It forms a quadrangle with
no chords if and only if the $(i, l)$ edge exists while the $(i, k)$
and $(j, l)$ edges do not. Since all edges in the ER model exist
independently with probability $p$ it means that the expected value of
global complementarity coefficient is $\mathbb{E}[c] = p(1-p)^2$.
Crucially, this result implies that global complementarity decays asymptotically
towards 0 in sparse random graphs ($\lim_{n \to \infty} \mean{d_i}/n \to 0$). 
This distinguishes it from global bipartivity measures
which attain non-minimal values for ER random graphs 
(cf.~Fig.~3a in Ref.~\cite{holmeNetworkBipartivity2003}).

\subsubsection*{Configuration model}

A classical null model for studying nodewise coefficients and their correlations
with node degrees is the configuration model in which a particular degree
sequence is enforced while apart from that connections are established as
randomly as possible~\cite{newmanNetworksIntroduction2010}.
In order to describe the qualitative behavior of the nodewise structural
similarity and complementarity we will use the fact that in both cases
they are bounded by their corresponding clustering and closure coefficients.

First, note that it is usually conjectured that $t$-clustering
should generally decrease with node degree~\cite{newmanNetworksIntroduction2010}.
More recently, it was analytically proven for the family of random networks
with power law degree distributions that $t$-clustering is on average roughly
constant for low-degree nodes and then starts to decrease more quickly
as node degree grows~\cite{vanderhofstadTriadicClosureConfiguration2018}.

On the other hand, it has been shown that local closure coefficient,
or $t$-closure in our terminology, is positively correlated
with node degree in the configuration model~\cite{yinLocalClosureCoefficient2019}.
Thus, these two results together imply that structural similarity $s_i$
can display rich, also non-monotonic, correlations with
node degrees depending on the structure of a particular network.

We leave analytical study of the analogous properties of $q$-clustering
and $q$-closure for future work. However, since both types of clustering
and closure coefficients are based on either wedge or head triples/quadruples
and therefore are very similar by construction, we conjecture that they should
display the same qualitative behavior in the configuration model.
Namely, we expect that $q$-clustering should decrease with node degree,
especially for well-connected nodes, and $q$-closure should increase with node
degree. As a result, we also expect that structural complementarity should vary
with respect to node degree in various, also non-monotonic, ways.

\begin{figure}[th]
\centering
\includegraphics[width=\textwidth]{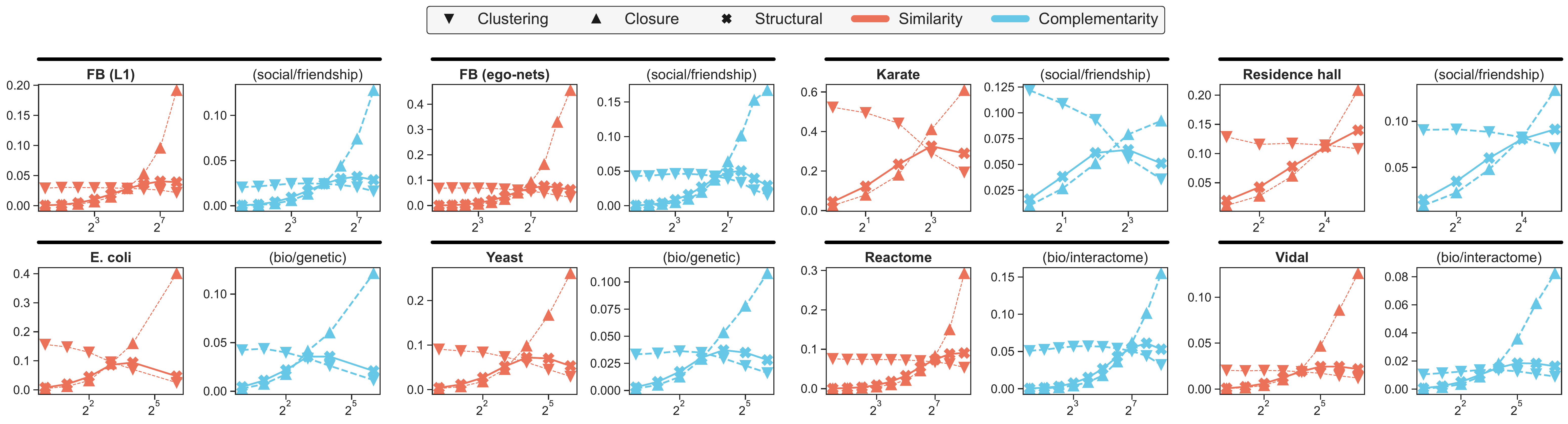}
\caption{%
Correlations of clustering, closure and structural coefficients with
node degrees in configuration model ensembles based on degree sequences
from 8 real-world social and biological networks.
See the SI (Fig.~S1) for full results based on 28 networks
and the description of the datasets.
Null distributions of the coefficients were approximated based on 100 samples 
from Undirected Binary Configuration Model (UBCM). The plots show values averaged for
different node degrees in logarithmic bins (base 2). As evident in the
figure, structural (similarity and complementarity) coefficients are always
bounded between their corresponding clustering and closure coefficients.
Furthermore, in all networks clustering coefficients tend to decrease
for high degree nodes while closure coefficients grow with respect to
degree. 
}
\label{fig:degrees}
\end{figure}

Indeed, our theoretical expectations agree with average trends observed in
randomized networks sampled from Undirected Binary Configuration
Model\cite{vallaranoFastScalableLikelihood2021} (UBCM; see Materials and Methods)
fitted to degree sequences of 28 real-world networks.
See Fig.~\ref{fig:degrees} for details. 
The results have two important practical implications. Firstly, structural coefficients
often tend to follow closure coefficients more closely for low-degree nodes and clustering
coefficients for high degree nodes. In other words, in the configuration model local structure
around low-degree (high-degree) nodes is dominated by head (wedge) triples/quadruples, that is,
clustering/closure coefficients are good descriptors of the density of triangles/quadrangles
only for particular subsets of the degree spectrum. More generally, the degree to which
they are relevant depends on the relative 
abundances of wedge and head paths. On the other hand, structural coefficients are more universal 
since they are weighted averages of both clustering and closure coefficients
with weights reflecting the relative dominance of wedge or head paths.

Secondly, structural coefficients depend on node degrees even in random graphs 
and therefore, when comparing different networks, their values should be calibrated 
based on a plausible null model such as UBCM to account for the effects induced 
purely by the first-order structure (degree sequences).

\subsection*{Structural coefficients in real networks}

\begin{figure}
\centering
\includegraphics[width=.8\textwidth]{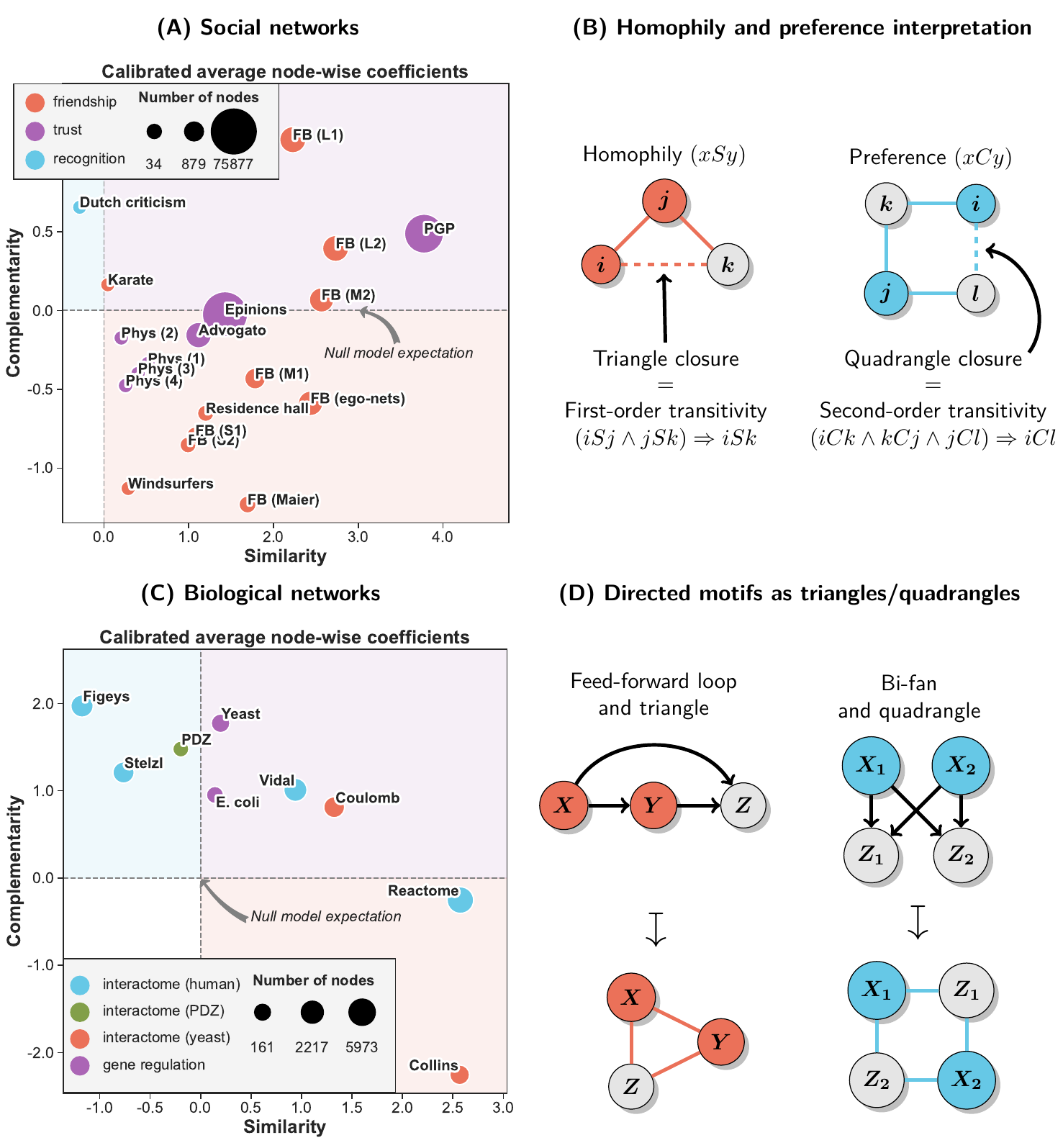}
\caption{%
Structural similarity and complementarity in social
(19 cases) and biological (9 cases) networks.
Scatterplots show calibrated average nodewise coefficients
with dashed lines indicating null model expectations based on UBCM
(see Materials and Methods for details on the datasets and calibration).
Colors of the quadrants of the plots indicate configurations of
increased values of similarity and complementarity coefficients ---
high and low (red), low and high (blue), both high (violet).
\textbf{(A)}~Social networks. Almost all feature high similarity 
and some also increased complementarity
(these are mostly large online social networks). The only case with low
similarity and relatively high complementarity is the network of Dutch
literary criticism representing relationships of recognition
(mentioning other's work, positively or negatively, in an essay or an interview)
within a set of notable literary authors~\cite{denooyLiteraryPlaygroundLiterary1999}.
\textbf{(B)}~Interpretation of similarity and complementarity in terms
of homophily, preferences and transitivity. Some social relations, especially those
depending on close bonds such as friendship or trust, are often driven by
homophily~\cite{marsdenHomogeneityConfidingRelations1988,mcphersonBirdsFeatherHomophily2001}.
This implies transitivity of ties and the abundance of triangles
due to triangle closure. However, other relations such as recognition or
skill-based collaboration~\cite{xieSkillComplementarityEnhances2016}
are based on preferences decoupled from the properties of the ego.
In this case two nodes with similar preferences connect to the same
neighbors but not necessarily to each other. This leads to what can be
called second-order transitivity, which in turn implies quadrangle closure.
\textbf{(C)}~Biological networks. Most of them feature both
increased similarity and complementarity indicating higher structural
diversity than in the case of social networks. This is consistent with
multiple results reporting the abundance of both triangle
and quadrangle based motifs~\cite{miloNetworkMotifsSimple2002,shen-orrNetworkMotifsTranscriptional2002,tranCountingMotifsHuman2013}.
\textbf{(D)}~Examples of the relationship between directed motifs often
reported for biological networks and undirected motifs used for defining 
structural similarity and complementarity.
}
\label{fig:domains}
\end{figure}

We studied structural similarity and complementarity in multiple real-world
social and biological networks measuring different kinds
of relations --- friendship, trust and recognition for social networks
as well as gene transcription regulation and general protein-protein
interactions (interactomes) for biological networks
(see Fig.~\ref{fig:domains} for details).
The goal was to see whether structural similarity and complementarity can be
related to some meaningful domain-specific properties of different types of
networks.

Our results show that similarity and complementarity in social networks
are indeed related to different types of relations. In particular,
similarity is stronger in systems driven by homophily, that is, 
preference for connecting to others who are similar to us,
which leads to the transitivity of relations. 
The importance of similarity seems to be particularly strong for
relations depending on close ties such as friendship or trust. This is
consistent with decades of research on social
networks~\cite{marsdenHomogeneityConfidingRelations1988,mcphersonBirdsFeatherHomophily2001,kossinetsOriginsHomophilyEvolving2009,richtersTrustTransitivitySocial2011}.
On the other hand, it seems that complementarity plays an important role in
shaping of relations in which preferences are decoupled from the properties
of the ego, such as recognition (e.g. of value or importance of others),
skill-based collaboration~\cite{xieSkillComplementarityEnhances2016}
or trade/business interactions~\cite{mattssonFunctionalStructureProduction2021}.
In this case two agents with similar preferences should typically connect
to the same neighbors (and therefore be structurally equivalent) but not
necessarily to each other, as the preferences of an agent do not have to
match its intrinsic properties. This leads to the
abundance of quadrangles and the presence of locally dense bipartite-like
subgraphs, that is, the structural signatures of complementarity.
Interestingly, even though such preference-based relations are not directly
transitive, they can be considered second-order transitive due to the implied
mechanism of quadrangle closure (see~Fig.~\ref{fig:domains}B). We put this
tentative hypothesis to a more direct and systematic test in the next section
(Similarity and complementarity in social relations).

Most of the biological networks feature both relatively high similarity and
complementarity. This is consistent with multiple results concerning network
motifs characteristic for interactomes as well as neural and gene transcription
regulatory networks~\cite{miloNetworkMotifsSimple2002,shen-orrNetworkMotifsTranscriptional2002,tranCountingMotifsHuman2013}.
Namely, structural similarity is linked to the presence of feed-back and
feed-forward loops which, when edge directions are unknown or ignored, explains
the abundance of triangles. On the other hand, structural complementarity is
connected to motifs such as bi-fan and bi-parallel~\cite{miloNetworkMotifsSimple2002},
which imply the abundance of quadrangles (see Fig.~\ref{fig:domains}D).
Importantly, these structural patterns can be linked to meaningful
domain-specific complementarities between different subsets of elements of a
system. For instance, in gene transcription regulatory networks bipartite-like
subgraphs with high density of bi-fan motifs (quadrangles) represent dense
overlapping regulons (DOR) or groups of operons regulated by similar
combinations of input transcription factors~\cite{shen-orrNetworkMotifsTranscriptional2002}.

Our results also point to important differences between social
and biological networks. The former, with some exceptions of course,
tend to be dominated by similarity while the latter are more
structurally diverse, which probably reflects their heterogeneous
functional properties and complex evolutionary history
(we study this in more detail in Section: Structural diversity across the tree of life). 
However, it seems that large online social networks also feature increased
complementarity relatively often (see Fig.~\ref{fig:domains}A).
Thus, it may be worthwhile to study differences between small and large
as well as offline and online social networks in the future. In particular,
to our best knowledge it is not yet clear what social processes are responsible
for significantly high amounts of quadrangles in large online social networks.

\subsection*{Similarity and complementarity in social relations}

Here we test the hypothesis that social relations based on homophily
are linked to structural similarity and those based on preference, 
recognition or skill-based collaboration to structural complementarity.
In other words, here we assess the theoretical validity of our approach.
For this purpose, we used a set of 34 social networks
collected in 17 rural villages in Mayuge District,
Uganda~\cite{chamiSocialNetworkFragmentation2017}.
For each village two networks of relations between households were measured:
(1)~a~friendship network and (2)~a~health advice network
(see Materials and Methods for details).

This dataset has the structure of a natural experiment as for each village
we have two different networks
representing relations between the same
households in the same period of time which were measured by the same research team(s)
using the same method. Thus, they are very likely to be equivalent with respect to any
possible covariate except for the type of relation that was measured 
(friendship or health advice). In other words, they can be compared to each other
as nearly perfect synthetic controls~\cite{craigNaturalExperimentsOverview2017}
and therefore allow reliable estimation of the effects specific for friendship
and health advice relations.

Thus, the dataset provides a perfect setting for
testing our hypothesis. Namely, it is sociologically justified to expect
the friendship networks to feature high structural similarity as it is a well
documented fact that friendship relations are to a large extent shaped by
homophily~\cite{marsdenHomogeneityConfidingRelations1988,mcphersonBirdsFeatherHomophily2001,kossinetsOriginsHomophilyEvolving2009}.
On the other hand, health advice networks should be at least partially driven
by complementarity, as the act of advice is usually based on the recognition
of and preference for one's knowledge as well as an information differential
between an adviser and an advisee. In other words, advising is based on a
synergy between needs and assets of two agents. Moreover, it can be also seen
as a particular kind of skill-based collaboration, which is known to be linked
to complementarity and
heterophily~\cite{riveraDynamicsDyadsSocial2010,xieSkillComplementarityEnhances2016}.
Thus, it is justified to expect the health advice networks to feature high
structural complementarity.

\begin{figure}[t]
\centering
\includegraphics[width=.7\linewidth]{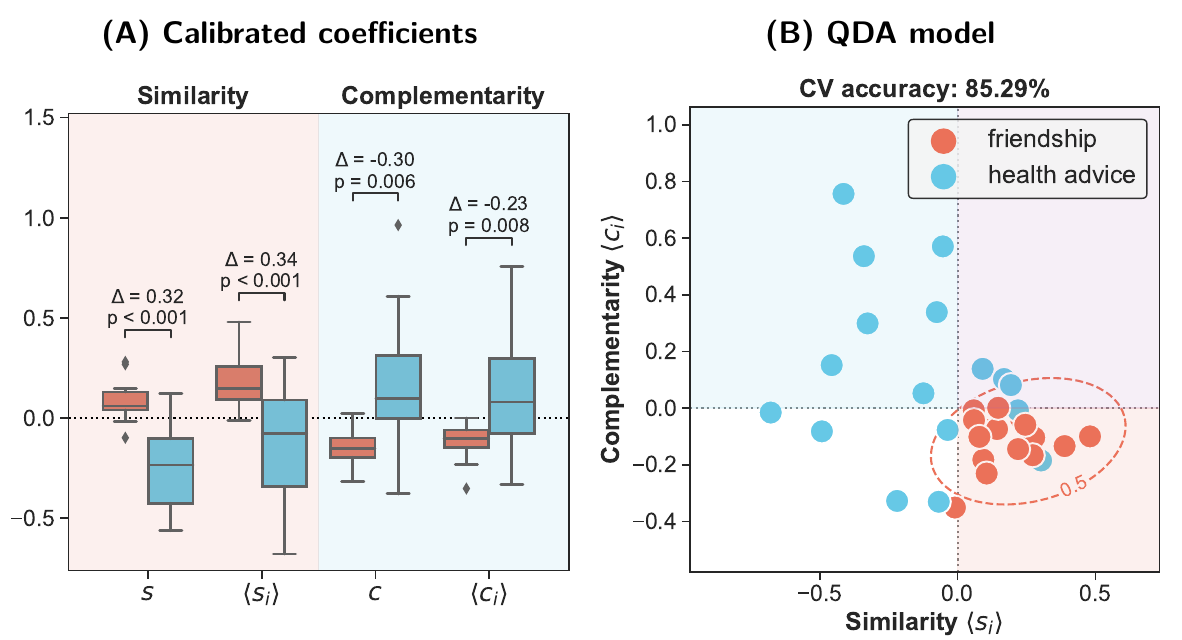}
\caption{%
Comparison of structural coefficients between friendship and
health advice networks in 17 Ugandan villages.
Observed values were calibrated based on 500 samples from UBCM
(see Materials and Methods).
\textbf{(A)}~Univariate distributions of global and average nodewise
coefficients reveal significant differences between the friendship and
health advice networks which are consistent with the hypothesis.
The friendship networks feature significantly higher structural similarity
and the health advice networks higher complementarity. The average
differences between networks from the same villages are denoted by
$\Delta$'s. Statistical significance was assessed using one-sample $t$-test
with two-sided null hypothesis applied to the differences; 
$p$-values were adjusted for multiple testing using Holm-Bonferroni method.
\textbf{(B)}~Bivariate distribution of the calibrated values of
the average nodewise similarity and complementarity coefficients.
The decision boundary separating the friendship and health advice networks
(marked in red) is based on Quadratic Discriminant Analysis (QDA).
The out-of-sample classification accuracy was estimated with stratified
17-fold cross-validation (one fold per village).
}
\label{fig:social}
\end{figure}

As evident in Fig.~\ref{fig:social}A, the results are in clear agreement with
the theoretical expectations. The calibrated similarity coefficients
(see Materials and Methods)
in the friendship networks were typically increased relative to the null model
(average log-ratios greater than zero) and significantly higher than in the
health advice networks ($p < 0.001$). On the other hand, the results for
the complementarity coefficients were exactly opposite and in this case the
health advice networks featured significantly larger calibrated values
($p < 0.01$).

Thanks to the convenient quasi-experimental structure of the dataset and the
calibration accounting for differences in degree sequences the results provide
strong support for the claim that, \textit{ceteris paribus}, social relations
based on similarity and complementarity leave distinct structural signatures
in social networks which can be detected using structural coefficients.
In other words, we showed that, all else being equal, similarity-based ties
are linked to the abundance of triangles and those based on complementarities
to the abundance of quadrangles. This confirms the theoretical validity of
the proposed framework and shows that patterns captured by structural
coefficients are indeed related to meaningful domain-specific phenomena.
Crucially, it also shows that there are types of social relations which are
driven not by similarity but complementarity, so the default assumption of
homophily is not always adequate.

To gauge the discriminatory power of the coefficients better,
we fitted a supervised classifier based on Quadratic Discriminant
Analysis (QDA)~\cite{hastieElementsStatisticalLearning2008}.
To facilitate visualization we used only two predictors:
average nodewise similarity and complementarity coefficients.
The estimated out-of-sample accuracy was $85.29\%$ (Fig.~\ref{fig:social}B),
which provides further confirmation of the theoretical validity of our approach.

\subsection*{Structural diversity across the tree of life}

Functioning of all biological organisms depends on protein-protein interactions
(PPIs), which themselves are constrained by the presence of
compatible binding sites~\cite{kovacsNetworkbasedPredictionProtein2019}.
Hence, it can be argued that it is not similar but complementary proteins that
are most likely to interact, or that two proteins sharing a neighbor do not
have to be connected but instead are likely to share other neighbors
(and be structurally equivalent). This view is supported by the statistical
over-representation of quadrangle-based motifs in interactome
networks~\cite{miloNetworkMotifsSimple2002,shen-orrNetworkMotifsTranscriptional2002}
as well as recent advances in PPI prediction, which showed that models based
on 3-paths (L3) and quadrangle closure outperform those based on 2-paths (L2)
and triangle closure~\cite{kovacsNetworkbasedPredictionProtein2019}. Moreover,
there is substantial evidence that protein neighborhoods in interactome
networks across the tree of life tend to gradually shift from the dominance
of triangles to quadrangles during
evolution~\cite{zitnikEvolutionResilienceProtein2019}.
Nonetheless, triangle-based motifs are also prevalent in PPI networks and
their presence tend to even correlate positively with the abundance of
quadrangles~\cite{tranCountingMotifsHuman2013}. Here we study this problem
from the perspective of structural similarity and complementarity and show
that increasing complexity of organisms is associated with higher structural
diversity of PPI networks, meaning that protein neighborhoods tend to feature
increasing numbers of both triangles and quadrangles.

\begin{figure}
\centering
\includegraphics[width=.75\textwidth]{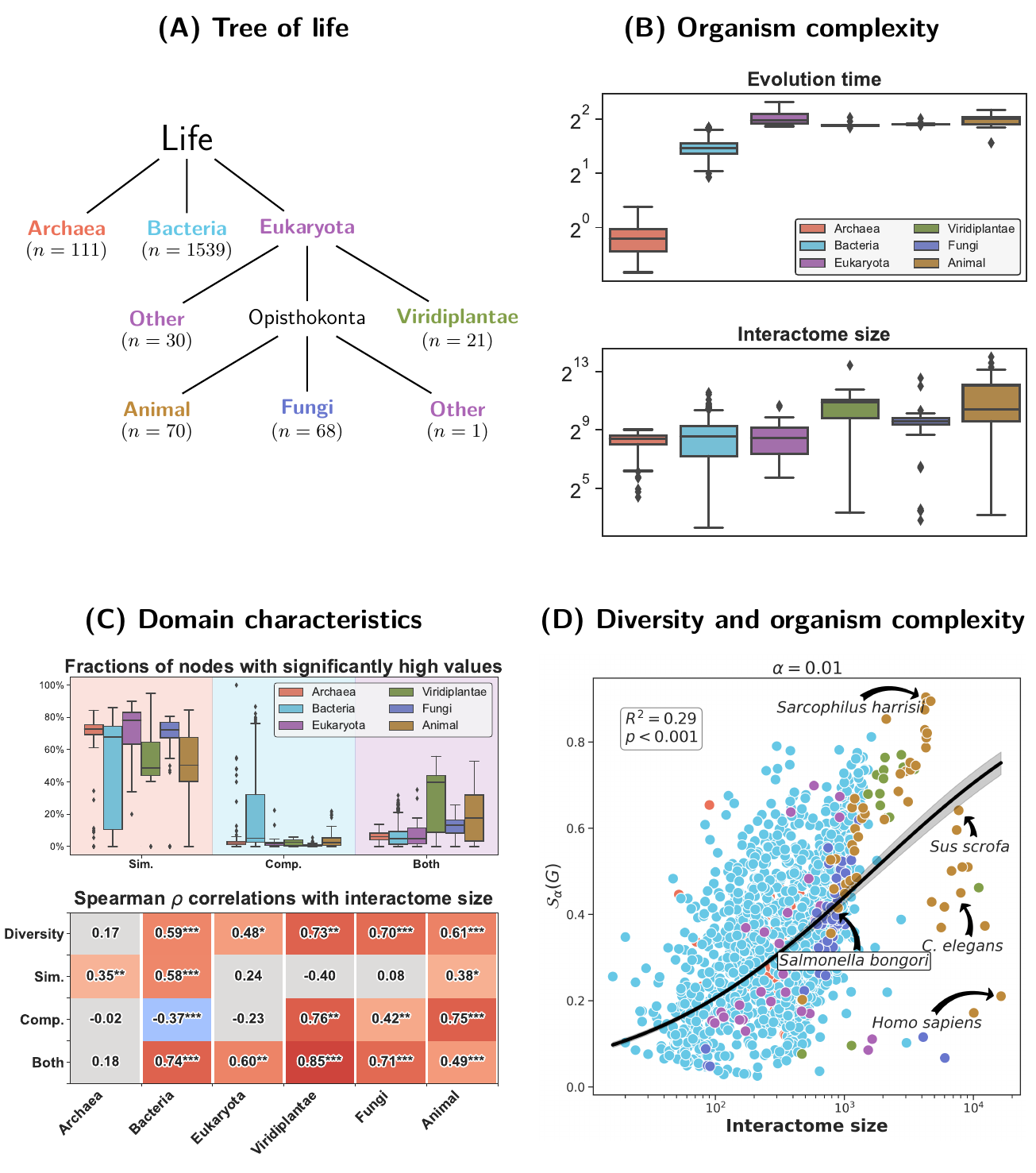}
\caption{%
Structural diversity of 1840 interactomes across the tree of
life~\cite{zitnikEvolutionResilienceProtein2019} (see Materials and Methods).
The analysis was based on proportions of
nodes with significantly high values ($p \leq \alpha = 0.01$) of structural
coefficients $s_i$ and $c_i$ or both of them. We also conducted
a sensitivity analysis for alternative levels $\alpha = 0.05, 0.10$
(SI: Structural diversity analysis).
Significance was based on null distributions estimated with 100
samples from UBCM. 
\textbf{(A)}~We followed the division into the domains of Archaea,
Bacteria and Eukaryota~\cite{woeseNaturalSystemOrganisms1990},
but also distinguished three arguably most complex taxa within
eukaryotes: green plants (Viridiplantae), fungi and animals (Metazoa).
The two \enquote{other} classes consist of all other eukaryotes.
\textbf{(B)}~Evolution time and interactome sizes in groups.
\textbf{(C)}~Proportions of nodes with significant
values of $s_i$ are high in all groups, but tend to be lower among bacteria,
which are the only group that feature relatively large
numbers of proteins with high structural complementarity.
Crucially, it is the more complex eukaryotes, particularly animals and plants,
which feature many nodes with significantly high values of both $s_i$ and $c_i$.
This suggests that interactomes of more complex organisms tend to be more
structurally diverse. The lower panel shows Spearman $\rho$ correlations
between interactome size and the proportions as well as the diversity
index (\ref{eq:diversity}); $^{***}p \leq 0.001$,${ }^{**}p \leq 0.01$,
${ }^{*}p \leq 0.05;$ Holm-Bonferroni correction for multiple testing was used.
Notably, in all groups except Archaea
the correlations between interactome size and the diversity index as well
as the \enquote{both} proportion are strongly positive. Moreover,
complementarity increases with network size in plants, fungi and animals
but decreases in bacteria, which means that in both cases the general trend
goes in the direction of greater structural diversity.
\textbf{(D)}~Structural diversity and interactome size.
The general correlation is positive and relatively strong.
}
\label{fig:proteins}
\end{figure}

We studied PPI networks, or interactomes, of 1840 species across the tree of
life~\cite{zitnikEvolutionResilienceProtein2019}
(see Fig.~\ref{fig:proteins} for details). We used network size
(number of proteins) for a proxy of the biological complexity of an organism,
which is arguably justified as on average interactomes of more complex organisms,
such as animals or green plants, are markedly larger than those of bacteria or
archaea. Moreover, taxa with larger interactomes on average also
tend to have longer average evolution times 
measured in terms of nucleotide substitutions per site
(Fig.~\ref{fig:proteins}B).

The analysis was focused on the structural diversity of protein neighborhoods
in terms of the local abundance of triangles and quadrangles in relation to
the organism complexity (interactome size). We quantified the structure at
the level of entire networks in terms of fractions of nodes with significantly
high values of $s_i$ and $c_i$ coefficients or both of them
(see Fig.~\ref{fig:proteins} for details). Moreover, we also combined the
fractions in a synthetic index of structural diversity,
$\mathbb{S}_{\alpha}(G) \in [0, 1]$
(see Materials and Methods for details on calculating $p$-values and structural diversity).

Our analysis (see Fig.~\ref{fig:proteins} for details) indicates a large
amount of variation between different species and taxa.
It suggests that bacteria interactomes tend to be
driven by complementarity, and therefore dominated by quadrangles, to a larger
extent than those of other organisms. On the other hand, more complex eukaryotes
(green plants, fungi and animals) tend to feature nodes with both high
structural similarity and complementarity more often, which implies that
protein neighborhoods in their interactomes are more heterogeneous and
contain both many triangles and quadrangles. Crucially, this intuition is
also confirmed by our structural diversity index which correlates positively
with organism complexity (interactome size) (Fig.~\ref{fig:proteins}D). 
Apart from the tail composed of species with large
PPI networks where the trend seems to bifurcate into two groups of organisms
with unexpectedly high and low diversity scores
(with some notable outliers such as \textit{Homo sapiens} and
\textit{Sarcophilus harrisii}, or Tasmanian devil), the model provides
a relatively good representation of the data generating process.
We modeled the relationship using a linear model with
logit transform applied to the diversity index and log transform to the
number of nodes. Thus, the relationship between \enquote{odds}
of the diversity index and the number of nodes follows a power law,
$\mathbb{S}_\alpha(G) / (1-\mathbb{S}_\alpha(G)) \propto n^{\gamma}$,
with $\gamma = 0.48$ (95\% CI: $[0.45, 0.51]$; $p < 0.001$).
We discuss additional details and analyses in the SI (Structural diversity analysis).
In particular, we study the stability of the results for different choices of $\alpha$
and examine models controlling for the number of publications on different species
(to partially correct for publication bias and resulting differences in terms of
interactome completeness).

The results suggest a general tendency towards greater structural diversity in
PPI networks of more complex organisms. In many cases this implies
an increasing prevalence of quadrangles, which is consistent with the results
reported in Ref.~\cite{zitnikEvolutionResilienceProtein2019}
as well as the general importance of complementarity of binding sites for
protein-protein interactions~\cite{kovacsNetworkbasedPredictionProtein2019}.
It is also consistent with the accounts of gene duplication occurring during evolution, 
and in particular whole genome 
duplication events~\cite{wolfeMolecularEvidenceAncient1997,dehalTwoRoundsWhole2005}, 
resulting in the creation of pairs of similarly wired proteins, 
which together may form multiple quadrangles.
These are tentative results which needs to be corroborated with more in-depth analyses
before they could have a substantial biological interpretation. Nonetheless, the general
picture painted by structural coefficients seems to agree with the
existing literature on PPI networks, which suggests that the proposed coefficient 
may be useful for studying biological networks.

Our results also indicate that, 
despite the likely increasing importance of quadrangles during evolution,
triangles are still important,
perhaps as a manifestation of feed-back and feed-forward loops,
and interactomes often feature many triangles and quadrangles at the same time,
which is consistent with the reports of positive correlations between triangle and quadrangle
densities in interactomes~\cite{tranCountingMotifsHuman2013}.
This suggests a way for improving on PPI prediction models based purely
on L2 or L3~\cite{kovacsNetworkbasedPredictionProtein2019}
measures by using a model averaging combining the two metrics
by somehow using the information on the local structure provided by
structural coefficients. We leave a detailed exploration of this idea for
future work.

\section*{Discussion}

Starting from first principles based on simple geometric arguments we
introduced a framework for measuring similarity- and complementarity-driven
relations in networks. We linked both relational principles to their characteristic
network motifs --- triangles and quadrangles respectively --- and defined
two general families of structural similarity and complementarity coefficients
measuring the extent to which they shape the structure of any
unweighted and undirected network.
In other words, we showed that both similarity and complementarity leave
statistically detectable structural signatures, which opens up new
possibilities for studying the structure of various networked systems
explicitly in terms of the impact of these two relational principles.
We also demonstrated, using multiple empirical examples, that both similarity and
complementarity are important for many kinds of social and biological relations.
In particular, our results indicate that the customary assumption of homophily may
not be appropriate for some social networks, of which structure may be better explained
by complementarity.

Even though the connection between the structure of networks and the principle of complementarity
is still relatively unexplored, our work was informed by existing studies
on quadrangle formation~\cite{jiaMeasuringQuadrangleFormation2021},
functional structure~\cite{mattssonFunctionalStructureProduction2021},
geometry of complementarity-driven networks~\cite{kitsakLatentGeometryComplementarityDriven2020}
and complementarity-based link prediction~\cite{kovacsNetworkbasedPredictionProtein2019}.
It extends this branch of the literature by introducing a set of general graph-theoretical
coefficients measuring the density of quadrangles and proposing a simple, minimalistic
geometric model linking the principle of complementarity to quadrangles as its
characteristic motif.

Furthermore, in contrast to previous studies using quadrangle-based descriptors of
local structure~\cite{jiaMeasuringQuadrangleFormation2021},
our approach is focused specifically on strong (chordless) quadrangles
(cf.~Fig.~\ref{fig:comp}A). This makes it, of course, less general, 
but at the same time allows making a direct connection between the principle
of complementarity and network bipartivity.
As a result, our work shows that the principle of complementarity induces
structures which are both locally bipartite-like and dense, in the same way
as similarity is connected to locally dense unipartite subgraphs. Moreover,
the proposed structural complementarity coefficients, which measure both bipartivity
and density, may be a useful addition to the existing set of measures of 
bipartivity~\cite{holmeNetworkBipartivity2003,estradaSpectralMeasuresBipartivity2005},
which do not consider local density. In particular, it may be potentially very useful
in studies on systems with so-called functional structure such as production/trade
or PPI networks, which are supposed to be characterized by both relatively high bipartivity 
and density of quadrangle motifs~\cite{mattssonFunctionalStructureProduction2021}.

Using structural coefficients applied to a rich empirical material,
we confirmed that typically social relations such as friendship or trust are
driven by similarity and therefore are transitive and linked to the abundance
of triangles. However, we also showed that some types of relations,
for instance advice, recognition or skill-based collaboration,
are more likely to be driven by complementarity,
which leads to markedly different local connectivity structures
dominated by quadrangles instead of triangles. Importantly,
this indicates that such relations are not directly transitive
($i \sim j \land j \sim k \Rightarrow i \sim k$),
but instead second-order transitive
($i \sim j \land j \sim k \land k \sim l \Rightarrow i \sim l$),
which implies that the principle of triangle (2-path) closure does not capture
the dynamics of such systems very well. Instead, it is quadrangle (3-path)
closure which is more adequate, so the default assumption of
homophily/triadic closure~\cite{mcphersonBirdsFeatherHomophily2001,kossinetsOriginsHomophilyEvolving2009,asikainenCumulativeEffectsTriadic2020}
is not always justified. Thus, our results encourage more nuanced approaches
to social network analysis and potentially can be used to design novel, more
flexible link prediction methods.

We also confirmed that biological networks such as gene transcription
regulatory or general PPI networks are more likely to be driven by
complementarity and feature more quadrangles than typical social networks.
This is consistent with multiple empirical results~\cite{shen-orrNetworkMotifsTranscriptional2002,tranCountingMotifsHuman2013,kovacsNetworkbasedPredictionProtein2019,zitnikEvolutionResilienceProtein2019}
and the general mechanism of protein-protein interactions based on
complementarity of binding sites~\cite{kovacsNetworkbasedPredictionProtein2019}.
Using structural coefficients, we demonstrated that interactome networks
of more complex organisms across the tree of life tend to be more structurally
diverse, meaning that they consist of many proteins with neighborhoods
containing significantly high numbers of both triangles and quadrangles.
This indicates a large degree of heterogeneity of structure in PPI networks
and suggests that recent results showing that protein interaction prediction
based on 3-path (L3) closure is more effective than the 2-path (L2)
closure rule~\cite{kovacsNetworkbasedPredictionProtein2019},
could be perhaps further improved by combining the L2 and L3 principles
in a way informed by the local structure around a given pair of proteins.

An important limitation of our work is the fact that our methods currently
can be applied only to undirected and unweighted networks. However,
generalizing them to the weighted case should be rather straightforward,
and we plan to address this problem in the future. In particular, it should
be possible to define weighted structural coefficients following the approach
used for defining weighted clustering coefficient in
Ref.~\cite{barratArchitectureComplexWeighted2004}.
On the other hand, the geometric motivation of structural coefficients
is inherently undirected, so it is not immediately clear how directed
coefficients should be defined. For now, we leave it as an interesting open
problem.

In summary, we showed that both similarity and complementarity are
important organizational principles shaping the structure of social and
biological networks and can be linked to interpretable,
domain-specific phenomena. We proposed a set of coefficients for measuring
the extent to which they shape the structure of networks and demonstrated the
theoretical validity and practical utility of the proposed framework on a rich
empirical material.

\section*{Materials and Methods}

\subsection*{Computing structural coefficients}

Structural coefficients are based on counting triples and triangles
(similarity) as well as quadruples and quadrangles (complementarity).
While the first problem is relatively easy and efficient methods for solving it
are implemented in many popular libraries for graph analysis,
the second problem of counting quadruples and quadrangles is more difficult
and corresponding efficient algorithms are not widely known.
Here we solve both problems by counting all motifs of interest
at the level of individual edges and then aggregate the edgewise counts
to nodewise or global counts when necessary. We propose an algorithm which
can be seen as a special case of a highly efficient exact graphlet counting
method proposed in Ref.~\cite{ahmedEfficientGraphletCounting2015}.
We call it \texttt{PathCensus} algorithm as ultimately it counts different types 
of paths and cycles. Pseudocode for the algorithm and other computational details 
are discussed in the SI (Structural coefficients and \texttt{PathCensus}).

\subsection*{Undirected Binary Configuration Model}

We used Undirected Binary Configuration Model
(UBCM)~\cite{vallaranoFastScalableLikelihood2021}
for the calibration and assessment of statistical significance of structural
coefficients. UBCM is a variant of the configuration model that induces
a maximum entropy probability distribution over undirected and unweighted
networks with $n$ nodes constrained to have a specific expected degree
sequence.

UBCM belongs to the family of Exponential Random Graph Models
(ERGM)~\cite{squartiniUnbiasedSamplingNetwork2015}
which induce maximum entropy distributions over networks satisfying
some constraints in expectation. Crucially, it means that such models
are fully specified by a set of sufficient
statistics~\cite{lehmannTheoryPointEstimation1998}
describing the desired constraints. Hence, the maximum entropy distributions
they induce are as unbiased as possible with respect to any other
property~\cite{squartiniUnbiasedSamplingNetwork2015}.

\subsection*{Calibrating values of structural coefficients}

In the analyses comparing different networks we calibrated observed values
of structural coefficients against UBCM in order to account for effects induced
purely by the first-order structure (i.e. degree sequences). Such a calibration
may be implemented in many different ways, but all reasonable approaches should
yield qualitatively comparable results. We explain our method using an
example of a calibration of a graph-level statistic such as average
nodewise similarity coefficient, $\mean{s_i}$.

First, for an observed network $G$ calculate the value of a graph statistic of
interest, $x(G)$. Then, sample $R$ randomized replicates $G_i$'s of the
observed network from a chosen null model (e.g.~UBCM) and calculate $x(G_i)$
for $i = 1, \ldots, R$. Finally, the calibrated value of $x(G)$
based on $R$ samples from the null model is defined as the
average log-ratio of the observed value and the randomized values:
\begin{equation}\label{eq:calibration}
    \mathcal{C}(x, R)(G)
    = \frac{1}{R}\sum_{i=1}^R \log{\frac{x(G)}{x(G_i)}}
\end{equation}

Note that the calibrated values are defined using ratios of $x(G)$ and $x(G_i)$'s,
which are expressed in the same units (e.g. triangles/2-paths), and therefore produce
a dimensionless quantity, as required by the logarithmic function~\cite{mattaCanOneTake2011}.

\subsection*{Assessing significance of structural coefficients}

Statistical significance of nodewise structural coefficients was estimated
using simulated null distributions based on $R$ samples from UBCM. We used
the fact that UBCM is a variant of the class of
ERGMs~\cite{vallaranoFastScalableLikelihood2021}
and therefore the probability distribution it induces is fully determined
by a set of sufficient statistics~\cite{lehmannTheoryPointEstimation1998},
that is, the expected degree sequence in our case. This implies that null
distributions of any statistics for nodes with the same degrees are identical,
so such nodes are indistinguishable from the vantage point of the model.
Thus, we estimated $p$-values according to the following procedure:
\begin{enumerate}
    \item Sample $R$ randomized analogues of an observed network $G$ from
    the probability distribution induced by UBCM.
    \item For each graph $G_i$ with $i = 1, \ldots R$ calculate a vector of
    nodewise statistics such as structural similarity coefficient $s_i$.
    \item Group calculated values in buckets defined by unique values of
    node degrees in the observed network $G$. Nodes in randomized networks
    are treated as if they had the same degrees as their corresponding nodes
    in $G$.
    \item Calculate quantiles of the distributions in the buckets.
    \item Set $p$-value for each node to $p = 1-\alpha_{\text{max}}$,
    where $\alpha_{\text{max}}$ is the maximum quantile lower than the observed
    value for a given node. In all cases we used one hundred quantiles or
    percentiles.
    \item Adjust $p$-values for multiple testing using two-stage
    False Discovery Rate (FDR) correction proposed by Benjamini, Krieger
    and Yekutieli (Definition~6 in Ref.~\cite{benjaminiAdaptiveLinearStepup2006}).
\end{enumerate}
Note that the above procedure ensures at least $R$ observations for each node
(and more for those with non-unique degrees) and therefore allows estimation
of $p$-values with a resolution of at least $0.01$ when $R \geq 100$ ($1/R$ in general).

\subsection*{Structural diversity index}

Let $p^\alpha_S(G), p^\alpha_C(G), p^\alpha_B(G)$ and $p^\alpha_N(G)$
be respectively proportions of nodes with significantly high values
(at $p \leq \alpha$) of $s_i$ or $c_i$ coefficients or both of them or neither
in a graph $G$. Then, we can define analogous proportions conditioned on the
set of nodes with at least one significant value as
$p^\alpha_{X \mid N'}(G) = p^\alpha_X(G) / (1 - p^\alpha_N(G))$
for $X = S, C, B$. The conditional proportions define a probability
distribution $\mathcal{P}^\alpha_G$. Finally, structural diversity index
of a graph $G$ at a significance level $\alpha$ is defined as:
\begin{equation}\label{eq:diversity}
    \mathbb{S}_\alpha(G) = \frac{(1 - p^\alpha_N)\mathbb{H}(\mathcal{P}^\alpha_G)}{\log_2{3}}
\end{equation}
where $\mathbb{H}(\mathcal{P}^\alpha_G) = -\sum_{X}p_X^\alpha(G)\log_2{p_X^\alpha(G)}$
is Shannon entropy functional~\cite{shannonMathematicalTheoryCommunication1948}
and $\log_2{3}$ term in the denominator is a normalizing constant
ensuring that $\mathbb{S}_\alpha(G) \in [0, 1]$.
This measure captures structural heterogeneity of node neighborhoods
while being penalized for networks with mostly random-like structure.

\subsection*{\texttt{pathcensus} package}

We implemented all the methods and algorithms for calculating structural
coefficients as well as several other utilities including most appropriate
null models and auxiliary methods for conducting statistical inference
in \texttt{pathcensus} package for Python. The core routines
are just-in-time compiled to highly optimized C code
using \textit{Numba} library~\cite{lamNumbaLLVMbasedPython2015} ensuring high
efficiency. The package has an extensive documentation including several usage
examples. It is currently available at GitHub (\url{https://github.com/sztal/pathcensus})
and will be distributed through \textit{Python Package Index} upon
publication.

\section*{Data availability}

This study did not generate any new data.
Networks used in this paper are freely accessible from the Netzschleuder repository:
\url{https://networks.skewed.de}.
Preprocessed data used in the analyses as well as the code needed for reproducing
the data and all the analyses are available at GitHub: 
\url{https://github.com/sztal/scs-paper} (the repository will be archived upon publication).

\section*{Acknowledgements}

We thank Shlomo Havlin for an advice on contextualizing our work
within the literature on network motifs as well as Brennan Klein and
Ivan Voitalov for an inspiring conversation on complementarity-driven relations
few years ago. We also thank Maciej Talaga for proofreading and
Mikołaj Biesaga for the help with testing the code.
This work was supported by a grant from National Science Center,
Poland (\textit{Outline of a network-geometric theory of social structure},
2020/37/N/HS6/00796).

\section*{Author contributions}

S.T. and A.N. conceptualized the project. S.T. formulated the mathematical formalism and wrote 
the related proofs, designed the algorithms and developed their Python implementation 
in the form of \texttt{pathcensus} package. S.T. conducted the data analyses and prepared 
the figures. S.T. and A.N. wrote the main text together.

\section*{Competing interests}

The authors declare no competing interests.

\section*{Additional information}

\textbf{Correspondence} and requests for materials should be addressed to S.T.

\bibliographystyle{ACM-Reference-Format}
\bibliography{SCS-SciRep.bib}
\addcontentsline{toc}{section}{References}

\FloatBarrier
\newpage

\titleformat{\section}%
    {\centering\normalfont\Large\bfseries}{\thesection.}{1em}{}
\renewcommand{\setthesubsection}{S\arabic{subsection}}

\setcounter{table}{0}
\renewcommand{\thetable}{S\arabic{table}}
\setcounter{figure}{0}
\renewcommand{\thefigure}{S\arabic{figure}}
\setcounter{equation}{0}
\renewcommand{\theequation}{S\arabic{equation}}
\setcounter{algorithm}{0}
\renewcommand{\thealgorithm}{S\arabic{algorithm}}

\section*{Supplementary Information}\label{app}
\addcontentsline{toc}{section}{Supplementary Information}

\begin{subappendices}

\section*{Similarity and structural equivalence}

Here we derive the relationship between similarity coefficient $s_{ij}$
and $s_i$ and structural equivalence.
First, we show that $s_i$ is a weighted average of the edgewise coefficients
$s_{ij}$'s for~$j \in \N_1(i)$, that is:
\begin{equation}\label{app:eq:sim-weighted-average}
    s_i =
    \frac{4T_i}{t^W_i + t^H_i} =
    \frac{\sum_j \left(t^W_{ij} + t^H_{ij}\right)s_{ij}}{\sum_j t^W_{ij} + t^H_{ij}}
\end{equation}
Note that Eq.~\eqref{eq:sim-edges} mplies that $\left(t^W_{ij} + t^H_{ij}\right)s_{ij} = 2T_{ij}$.
Moreover, since each triangle including $i$ is shared with two other neighbors we have that:
\begin{equation}\label{app:eq:sim-triangles}
    \sum_{j \in \N_1(i)}\left(t^W_{ij} + t^H_{ij}\right)s_{ij} = \sum_{j \in \N_1(i)} 2T_{ij} = 4T_i
\end{equation}
On the other hand, $t^W_{ij} + t^H_{ij}$ is the number of 2-paths traversing
the $(i, j)$ edges so it can be written as $t^W_{ij} + t^H_{ij} = d_i + d_j - 2$.
Hence, it is easy to see that:
\begin{equation}\label{app:eq:sim-triples}
\begin{split}
    \sum_{j \in \N_1(i)} t^W_{ij} + t^H_{ij}
    &= \sum_{j \in \N_1(i)}(d_i + d_j - 2) \\
    &= d_i(d_i-1) + \sum_{j \in \N_1(i)}(d_j - 1) \\
    &= t^W_i + t^H_i
\end{split}
\end{equation}
Finally, substituting \eqref{app:eq:sim-triangles} and \eqref{app:eq:sim-triples}
into \eqref{app:eq:sim-weighted-average} we confirm the desired equality.

Now, we use a common definition of structural equivalence in terms of
Sørenson Index (normalized Hamming similarity) and note its direct connection
to our notion of edgewise structural similarity $s_{ij}$:
\begin{equation}\label{app:eq:sim-edges-hamming}
    H_{ij}
    = \frac{2n_{ij}}{d_i + d_j}
    = \frac{2T_{ij}}{d_i + d_j}
    = s_{ij}\frac{d_i + d_j - 2}{d_i + d_j}
\end{equation}
The above implies that $H_{ij} < s_{ij}$ for all $(i, j)$ edges
for which $s_{ij}$ is defined. And since we established that $s_i$ is a
weighted average of $s_{ij}$'s with $j \in \N_1(i)$ we have that:
\begin{equation}\label{app:eq-sim:bounds}
    \min_j H_{ij} < \min_j s_{ij}
    \leq s_i \leq
    \max_j s_{ij} = \max_j\left(H_{ij}\frac{d_i + d_j}{d_i + d_j - 2}\right)
\end{equation}
Note that for large values of $d_i + d_j$ the above is approximately
equivalent to:
\begin{equation}\label{app:eq:sim-bounds-approx}
    \min_j H_{ij} < s_i \leq \max_j H_{ij}
\end{equation}

In other words, we showed that the similarity coefficient of a node $i$
is approximately bounded between minimum and maximum structural equivalence
(Sørenson Index) between itself and any of its neighbors.

\section*{Complementarity and structural equivalence}

Here we derive the relationship between complementarity coefficients $c_{ij}$
and $c_i$ and structural equivalence. We start by showing that
$c_i$ is a weighted average of the edgewise coefficients $c_{ij}$'s for
$j \in \N_1(i)$, that is:
\begin{equation}\label{app:eq:comp-weighted-average}
    c_{i} = \frac{4Q_{ij}}{q^W_i + q^H_i}
    = \frac{\sum_j \left(q^W_{ij} + q^H_{ij}\right)c_{ij}}{\sum_j q^W_{ij} + q^H_{ij}}
\end{equation}
Using Eq.~\eqref{eq:comp-edges} we can write $2Q_{ij} = \left(q^W_{ij} + q^H_{ij}\right)c_{ij}$.
Moreover, each strong quadrangle including a node $i$ is shared with exactly
two other neighbors. Hence, we have that:
\begin{equation}\label{app:eq:comp-quadrangles}
    \sum_{j \in \N_1} \left(q^W_{ij} + q^H_{ij}\right)c_{ij}
    = \sum_{j \in \N_1(i)} 2Q_{ij} = 4Q_i
\end{equation}
Next, note that each $3$-path starting at an
$(i, j)$ edge defines a unique ordered quadruple of the form $(i, j, k, l)$
or $(j, i, k, l)$. The first form is counted as a head quadruple of the node
$i$ and a wedge quadruple of the node $j$ and in the second case the order
is reversed. And since $q^W_{ij} + q^H_{ij}$ is the number of $3$-paths
starting at the $(i, j)$ edge it must hold that:
\begin{equation}\label{app:eq:comp-quadruples}
    \sum_{j \in \N_1(i)} \left(q^W_{ij} + q^H_{ij}\right) = q^W_i + q^H_i
\end{equation}
Finally, note that \eqref{app:eq:comp-quadrangles} and \eqref{app:eq:comp-quadruples}
jointly mean that \eqref{app:eq:comp-weighted-average} must be true.
As a result, for $j \in \N_1(i)$ we have that:
\begin{equation}\label{app:eq:comp-edge-node-bound}
    \min_j c_{ij} \leq c_i \leq \max_j c_{ij}
\end{equation}

Now, in order to derive the connection between complementarity coefficients
and structural equivalence we need first to introduce one additional quantity.
For a connected triple $(k, i, j)$ we define
Asymmetric Excess Sørenson Index:
\begin{equation}\label{app:eq:excess-sorenson-2}
    H_{kj|i} = \frac{n_{jk}-1}{d_k - 1 - a_{jk}}
\end{equation}
which measures how many of the connections of $k$ are also shared by $j$
while disregarding edges $(i, k)$, $(i, j)$ and $(j, k)$.

Next, we also need to use the notion of weak quadrangles
allowing for any number of chordal edges. Let $W_{ij} \geq Q_{ij}$ be the
number of quadrangles with any number of chords incident to the $(i, j)$ edge.
We also define weak edgewise complementarity to be
$h_{ij} = W_{ij} / (q^W_{ij} + q^H_{ij}) \geq c_{ij}$. It is easy to see that:
\begin{equation}\label{app:eq:Wij}
    W_{ij} = \sum_{k \in \N_1(i) - \{j\}} n_{jk} - 1
\end{equation}
On the other hand, the number of $3$-paths starting at the $(i, j)$ edge is:
\begin{equation}\label{app:eq:qij}
\begin{split}
    q^W_{ij} + q^H_{ij}
    &=
    \sum_{k \in \N_1(i) - \{j\}}(d_k-1) +
    \sum_{l \in \N_1(j) - \{i\}}(d_l-1) - 2n_{ij} \\
    &=
    \sum_{k \in \N_1(i) - \{j\}}(d_k-1 - a_{jk}) +
    \sum_{l \in \N_1(j) - \{i\}}(d_l-1 - a_{il})
\end{split}
\end{equation}
since $q^W_{ij}$ is the number of $(j, i, k, l)$ and $q^H_{ij}$ of
$(i, j, k, l)$ quadruples. The second equality comes from the fact that
$n_{ij} = \sum_k a_{jk} = \sum_l a_{il}$. Now, we can use
\eqref{app:eq:excess-sorenson}, \eqref{app:eq:Wij} and \eqref{app:eq:qij}
to rewrite the weak edgewise complementarity as:
\begin{equation}\label{app:eq:cij}
    h_{ij} =
    \frac{\sum_k (d_k-1 - a_{jk})H_{kj|i} + \sum_l (d_l-1 - a_{il})H_{li|j}}%
    {\sum_k (d_k-1 - a_{jk}) + \sum_l (d_l-1 - a_{il})}
\end{equation}
As a result, for $k \in \N_1(i)-\{j\}$ and $l \in \N_1(j)-\{i\}$ we have that:
\begin{equation}\label{app:eq:hij-bound}
    \min_{k,l}(H_{kj|i}, H_{li|j})
    \leq h_{ij} \leq
    \max_{k,l}(H_{kj|i}, H_{li|j})
\end{equation}
Using \eqref{app:eq:comp-edge-node-bound} we can write:
\begin{equation}\label{app:eq:comp-w2-bounds}
    \min_{j, k,l}(H_{kj|i}, H_{li|j})
    \leq h_{i} \leq
    \max_{j, k,l}(H_{kj|i}, H_{li|j})
\end{equation}
Finally, since by definition $c_{ij} \leq h_{ij}$ this implies:
\begin{equation}\label{app:eq:cij-bound}
    0
    \leq c_{ij} \leq
    \max_{k,l}(H_{kj|i}, H_{li|j})
\end{equation}
as well as:
\begin{equation}\label{app:eq:comp-bounds}
    0 \leq c_i \leq
    \max_{j, k, l}\left(H_{kj|i}, H_{li|j}\right)
\end{equation}

In other words, we just showed that $c_i$ is bounded from above by the maximum
Asymmetric Excess Sørenson Index between any two of its neighbors or itself
and any neighbor of its neighbors. Moreover, in the weak case we also have a lower
bound of the same nature. We leave a more detailed analysis of the notion
of weak complementarity for future work.

\section*{Structural coefficients and \texttt{PathCensus}}

\subsection*{Formulas and algorithm}

Edge-level counts of triples, quadruples, triangles and quadrangles
are computed with the algorithm~\ref{app:algo:pathcensus}.
Node and global counts can be obtained by aggregating edge counts.
The rules of aggregation are summarized in Table~\ref{app:tab:aggregation}.
Table~\ref{app:tab:formulas} presents detailed formulas for all
structural coefficients expressed in terms of the aggregated counts.

\begin{table}[htb!]
\centering
\caption{%
    Formulas for aggregating
    from edge to node and global counts
}
\label{app:tab:aggregation}
\begin{tabular}{cccc}
    \toprule
    & \multicolumn{3}{c}{Counting level} \\
    & Edge & Node & Global \\
    \midrule
    Paths \\
    \midrule
    Wedge triples & $t^W_{ij}$ & $t^W_i = \sum_{j}t^W_{ij}$ & $t^W = \frac{1}{2}\sum_{i,j}t^W_{ij}$ \\
    Head triples & $t^H_{ij}$ & $t^H_i = \sum_{j}t^H_{ij}$ & $t^H = \frac{1}{2}\sum_{i,j}t^H_{ij}$ \\
    Wedge quadruples & $q^W_{ij}$ & $q^W_i = \sum_{j}q^W_{ij}$ & $q^W = \frac{1}{2}\sum_{i,j}q^W_{ij}$ \\
    Head quadruples & $q^H_{ij}$ & $q^H_i = \sum_{j}q^H_{ij}$ & $q^H = \frac{1}{2}\sum_{i,j}q^H_{ij}$ \\
    \midrule
    Cycles \\
    \midrule
    Triangles & $T_{ij}$ & $T_{i} = \frac{1}{2}\sum_j T_{ij}$ & $T = \frac{1}{6}\sum_{i,j}T_{ij}$ \\
    Quadrangles & $Q_{ij}$ & $Q_i = \frac{1}{2}\sum_{j}Q_{ij}$ & $Q = \frac{1}{8}\sum_{i,j}Q_{ij}$ \\
    \bottomrule
\end{tabular}
\end{table}

\begin{table}[htb!]
\centering
\renewcommand{\arraystretch}{2}
\caption{%
    Formulas for structural coefficients based on path and cycle counts
}
\label{app:tab:formulas}
\begin{tabular}{cccc}
    \toprule
    & & \multicolumn{2}{c}{Relational principle} \\
    Level & Coefficient & Similarity & Complementarity \\
    \midrule
    Edges & Structural & $s_{ij} = \frac{2T_{ij}}{t^W_{ij} + t^H_{ij}}$ & $c_{ij} = \frac{2Q_{ij}}{q^W_{ij} + q^H_{ij}}$ \\
    \midrule
    Nodes & Structural & $s_i = \frac{4T_i}{t^W_i + t^H_i}$ & $c_i = \frac{4Q_i}{q^W_i + q^H_i}$ \\
          & Clustering & $s^W_i = \frac{2T_i}{t^W_i}$ & $c^W_i = \frac{2Q_i}{q^W_i}$ \\
          & Closure    & $s^H_i = \frac{2T_i}{t^H_i}$ & $c^H_i = \frac{2Q_i}{q^H_i}$ \\
    \midrule
    Global$^1$
          & Structural & $s = \frac{6T}{t^W + t^H}$ & $c = \frac{8Q}{q^W + q^H}$ \\
          & Clustering & $s^W = \frac{3T}{t^W}$ & $c^W = \frac{4Q}{q^W}$ \\
          & Closure    & $s^H = \frac{3T}{t^H}$ & $c^H = \frac{4Q}{q^H}$ \\
    \bottomrule
\end{tabular}

\begin{tablenotes}
    \centering
    \item $^1$All global measures are equivalent.
\end{tablenotes}
\end{table}

\FloatBarrier
\begin{algorithm}
\small
\caption{%
    \textbf{\texttt{PathCensus} algorithm.}
    It takes an undirected graph $G = (V, E)$ with $|V| = n$
    and $|E| = m$ as input and returns an array of edgewise counts of wedge
    and head triples and quadruples as well as triangles and (strong)
    quadrangles. For better performance $E$ can be defined
    (without loss of generality) to ensure that for all edges $(i, j)$
    it holds that $d_i \leq d_j$.
}
\label{app:algo:pathcensus}
\begin{algorithmic}[1]
    \State Initialize empty $C$ \Comment{$m \times 8$ array for storing path counts}
    \State Initialize $R$ such that $R_i = 0 \quad \forall i \in V$
    \Comment{$n \times 1$ array for keeping track of node roles}
    \State Let $D$ be the degree sequence of $G$ \Comment{$n \times 1$ array}
    \State Initialize $u = 0$

    \For{$e = (i, j) \in E$} \Comment{the loop may be parallelized}
        \State Set $u = u + 1$
        \State Initialize $T_{ij}, t^W_{ij}, t^H_{ij} = 0$
        \Comment{counts of triangles and wedge and head triples}
        \State Initialize $Q_{ij}, q^W_{ij}, q^H_{ij} = 0$
        \Comment{counts of strong quadrangles and wedge and head quadruples}

        \State Initialize $\text{Star}_i, \text{Star}_j, \text{Tri}_{ij} = \emptyset$
        \Comment{Empty sets for keeping track of nodes with different roles}

        \For{$(k \neq j) \in \N_1(i)$}
            \State Add $k$ to $\text{Star}_i$ and set $R_k = 1$
            \State Set $t^W_{ij} = t^W_{ij} + 1$
        \EndFor

        \For{$(k \neq i) \in \N_1(j)$}
            \If{$R_k = 1$}
                \State $T_{ij} = T_{ij} + 1$
                \State Remove $k$ from $\text{Star}_i$,
                add $k$ to $\text{Tri}_{ij}$ and set $R_k = 3$
            \Else
                \State Add $k$ to $\text{Star}_j$ and set $R_k = 2$
            \EndIf
            \State $t^H_{ij} = t^H_{ij} + 1$
        \EndFor

        \For{$k \in \text{Star}_i$}
            \Comment{This internal nested loop determines computational complexity}
            \For{$(l \neq i) \in \N_1(k)$}
                \If{$R_l = 2$}
                    \State $Q_{ij} = Q_{ij} + 1$
                \EndIf
            \EndFor
        \EndFor

        \For{$k \in \text{Star}_i$}
            \State Set $q^W_{ij} = q^W_{ij} + D_k - 1$ and $R_k = 0$
        \EndFor
        \For{$k \in \text{Star}_j$}
            \State Set $q^H_{ij} = q^H_{ij} + D_k - 1$ and $R_k = 0$
        \EndFor
        \For{$k \in \text{Tri}_{ij}$}
            \State Set $q^W_{ij} = q^W_{ij} + D_k - 2$
            \State Set $q^H_{ij} = q^H_{ij} + D_k - 2$
            \State Set $R_k = 0$
        \EndFor

        \State Set $C_u = \left(T_{ij}, t^W_{ij}, t^H_{ij}, Q_{ij}, q^W_{ij}, q^H_{ij}\right)$
        \Comment{Set $u$-th row of $C$}
    \EndFor
    \State \Return $C$
\end{algorithmic}
\end{algorithm}

\FloatBarrier
\subsection*{Calculating counts for reversed edges}

Note that in our implementation $t^W_{ij}$ counts the number of $(k, i, j)$
and $t^H_{ij}$ tracks $(i, j, k)$ triples. Thus, we have that
$t^W_{ij} = t^H_{ji}$. Similarly, $q^W_{ij}$ counts $(j, i, k, l)$
and $q^H_{ij}$ $(i, j, k, l)$ quadruples, so again we have that
$q^W_{ij} = q^H_{ji}$. On the other hand, counts of triangles and quadrangles
are symmetric. As a result, for the purpose of counting we can assume that all
edges are of the form $i < j$ and still be able to count everything correctly.
In other words there is no need to consider each undirected edge twice.

\subsection*{Computational complexity}

It is clear from the structure of the three nested loops that the asymptotic
worst-case computational complexity of both algorithms
is $O(m\Delta{}Sd_{\text{max}})$ where $m$ is the number of edges, $\Delta{}S$
is the maximum size of a $\text{Star}_i$ set
(see Algorithm~\ref{app:algo:pathcensus}) and $d_{\text{max}}$ is the maximum
node degree. This agrees with the analysis presented by the authors
of the more general graphlet counting
method~\cite{ahmedEfficientGraphletCounting2015},
which inspired our \texttt{PathCensus} algorithm. However, in practice the
runtime can be reduced by enforcing that edges are defined to satisfy
the condition $d_i \leq d_j$
(note that this can always be done without loss of generality).
The impact of this optimization can be quite significant for networks with
highly heterogeneous degree distributions. For instance, in the case of the
PGP web of trust network~\cite{richtersTrustTransitivitySocial2011}
($n = 39796, \mean{d_i} = 9.91, d_{\text{max}} = 1696$)
it yields almost 4 times shorter runtime on average.

\section*{Degree correlations in configuration model}
\FloatBarrier

\begin{figure}[htb!]
\centering
\includegraphics[width=.925\textwidth]{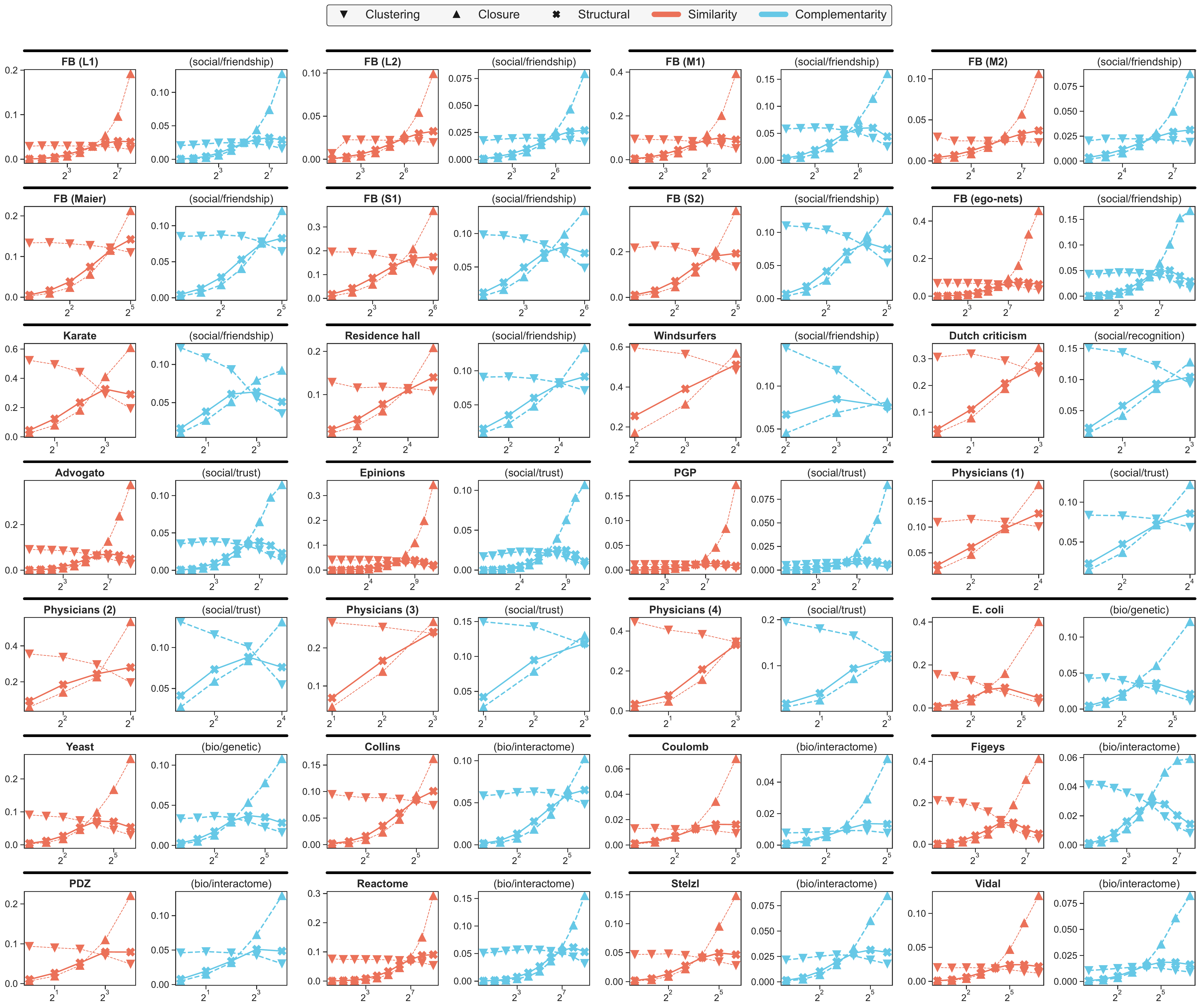}
\caption{%
    Correlations of clustering, closure and structural coefficients with
    node degrees in configuration model ensembles based on degree sequences
    from 28 real-world social and biological networks
    (see Section: Network datasets). Null distributions
    of the coefficients were approximated based on 100 samples from Undirected
    Binary Configuration Model (UBCM). The plots show values averaged for
    different node degrees in logarithmic bins (base 2).
}
\label{app:fig:degrees}
\end{figure}

\section*{Structural diversity analysis}

Here we provide additional details for the corresponding analysis in the
Main Text (Section: Structural diversity across the tree of life).
We present results for three different choices of significance level,
$\alpha = 0.01, 0.05, 0.10$ used for detecting nodes with high structural
similarity and complementarity. We show that qualitative results are stable
for all values of $\alpha$, even though quantitative details change in some
cases.

\begin{figure}[htb!]
\centering
\includegraphics[width=\textwidth]{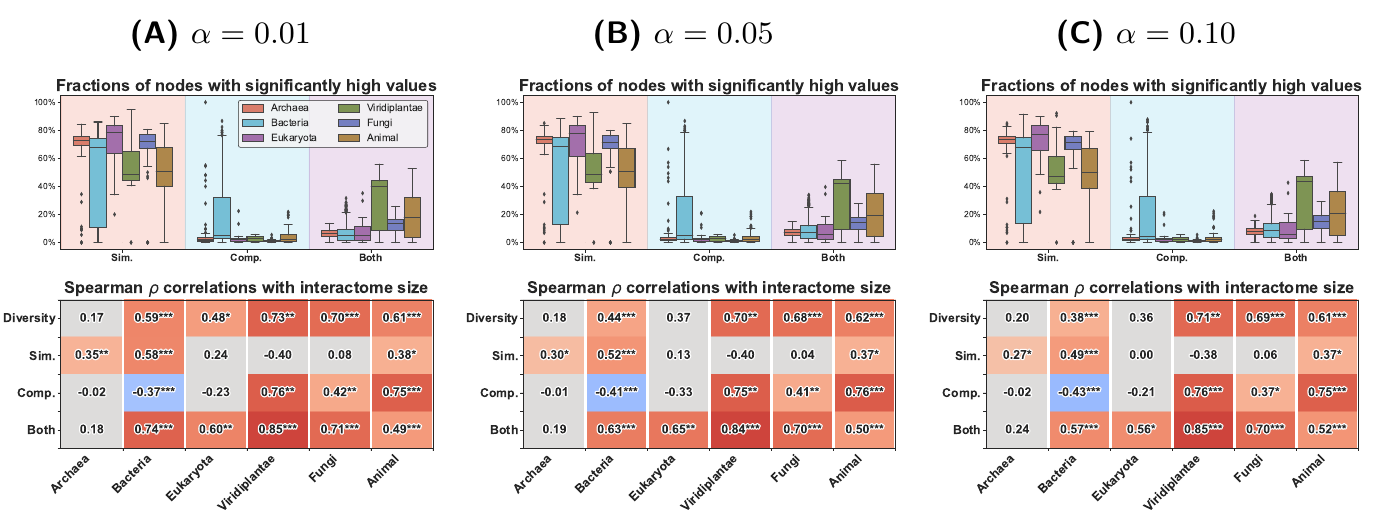}
\caption{
    General characteristic of the structure of interactomes in different
    domains/groups. Apart from some minor details the results for all values
    of $\alpha$ are the same.
}
\label{app:fig:differences}
\end{figure}

\subsection*{Linear model stability and diagnostics}

We described the relationship between structural diversity,
$y = \mathbb{S}_\alpha(G)$, defined in Eq.~(11) in the Main Text,
and interactome size $n$ (number of proteins) using a transformed linear model
of the form:
\begin{equation}\label{app:eq:transformed-lm}
    \eta(\hat{y}) = c + \gamma{}\log{n}
\end{equation}
where $\eta(x) = x / (1-x)$ is the logit transformation. This parametric form
ensured model predictions bounded in $(0, 1)$, which was necessary as the
structural diversity index ranges from $0$ to $1$. On the other hand, since
the logit transformation is not defined for $0$'s and $1$'s we had to drop
some part of the observations, namely, 119, 94 and 88 cases for
$\alpha = 0.01, 0.05, 0.10$ respectively. These observations corresponded
almost exclusively to organisms with small interactomes as indicated by
small average numbers of nodes ($35.07$, $30.98$ and $28.15$) as compared
to the overall average of $544.06$ nodes. Moreover, all of them were dropped
because of $\mathbb{S}_\alpha(G) = 0$, which is consistent with the hypothesis
that less complex organisms tend to have less structurally diverse interactomes.

\begin{figure}
\centering
\includegraphics[width=.9\textwidth]{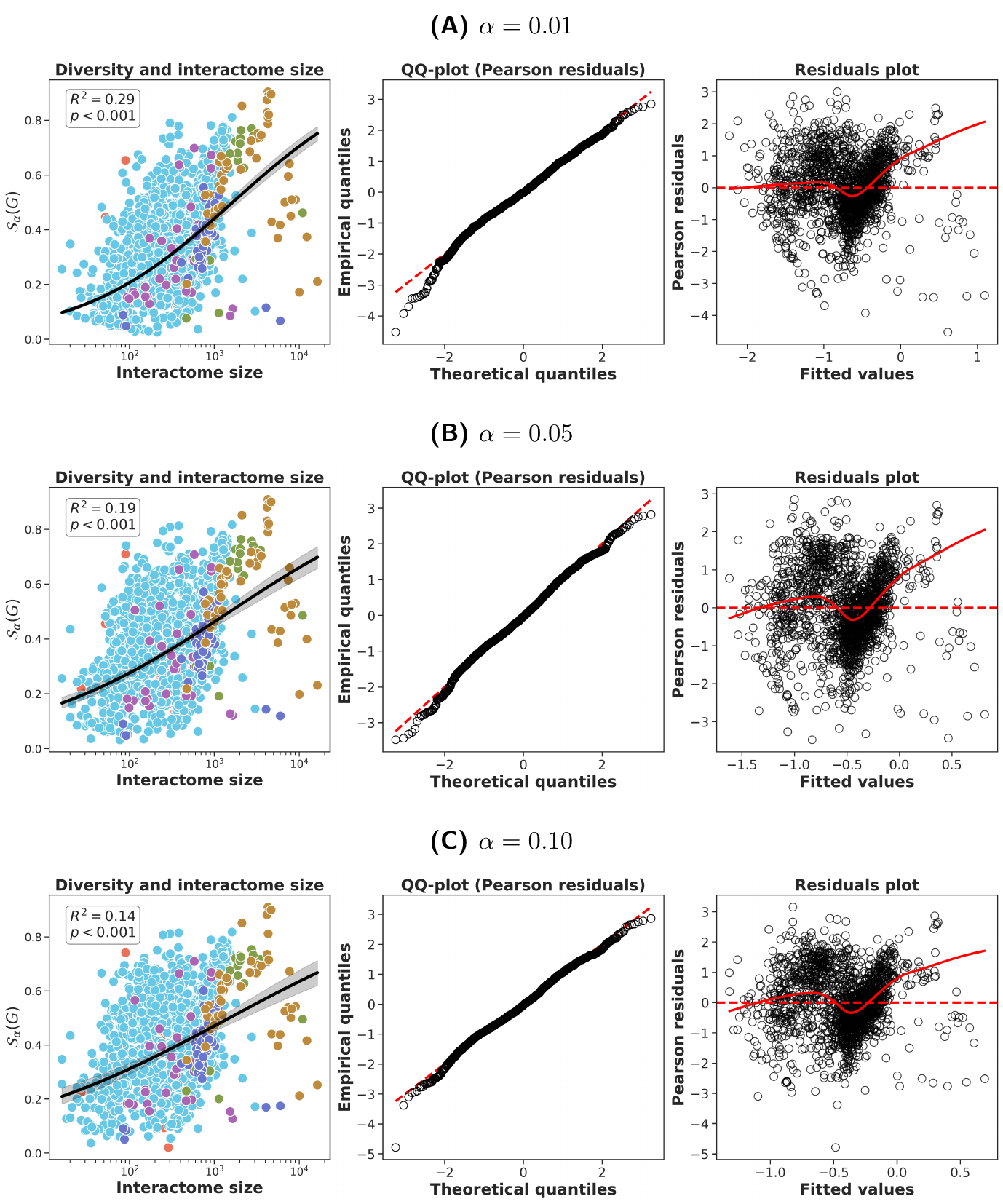}
\caption{
    Model fit and residual diagnostics for $\alpha = 0.01, 0.05, 0.10$.
    In all cases the qualitative trend is similar but $R^2$ drops for higher
    values of $\alpha$. Moreover, in all cases residuals are approximately
    normal apart from larger fluctuations in the region of high fitted values.
    This is an effect of the apparent bifurcation of the conditional
    distribution of structural diversity index for species with large
    interactomes discussed in the Main Text.
}
\label{app:fig:models}
\end{figure}

As evident in Fig.~\ref{app:fig:models} the qualitative trend is the same
for all values of $\alpha$. However, the goodness-of-fit of the model is
highest for $\alpha = 0.01$. This is not surprising. Higher values of $\alpha$
correspond to higher type I error rates, meaning that the estimated fractions
of nodes with significantly high values of $s_i$ and $c_i$ are more noisy.
Table~\ref{app:tab:lm} presents estimated parameters and other numerical
details.

Moreover, since the accuracy of interactome networks may depend on the extent
to which a given species has been studied, we also fitted extended models
including the logarithm of the number of publications about a given species
as the second predictor. 
This enabled a partial control for possible biases induced by differences
in terms of the incompleteness of the data available for more and less frequently
studied organisms.
As Table~\ref{app:tab:lm} shows, 
the effects of publication count ($b$) were insignificant in all cases.

\begin{table}[ht!]
\centering\small
\caption{Estimated parameters of the linear models$^1$}
\label{app:tab:lm}
\begin{tabular}{c|ccccc|ccccc|ccccc}
         & \multicolumn{5}{c}{$\alpha = 0.01$ $(N = 1721)$}
         & \multicolumn{5}{c}{$\alpha = 0.05$ $(N = 1746)$}
         & \multicolumn{5}{c}{$\alpha = 0.10$ $(N = 1752)$} \\
         \hline
         & Value & SE & $z$ & $p$ & & Value & SE & $z$ & $p$ & & Value & SE & $z$ & $p$ \\
         \hline
$c$      & -3.562 & 0.075 & -47.476 & < 0.001 & & -2.595 & 0.133 & -19.477 & < 0.001 & & -2.139 & 0.164 & -13.048 & < 0.001 & \\
$\gamma$ & 0.481 & 0.015 & 32.256 & < 0.001 & & 0.353 & 0.023 & 15.299 & < 0.001 & & 0.292 & 0.027 & 10.817 & < 0.001 & \\
\hline
         & \multicolumn{4}{l}{\scriptsize R$^2$ = 0.290, F(1719,1) = 1040, p < 0.001} &
         & \multicolumn{4}{l}{\scriptsize R$^2$ = 0.188, F(1744,1) = 234, p < 0.001} &
         & \multicolumn{4}{l}{\scriptsize R$^2$ = 0.290, F(1750,1) = 117, p < 0.001} & \\
         & \multicolumn{4}{l}{\scriptsize Skew = -0.296, Kurtosis = 3.728 (residuals)} &
         & \multicolumn{4}{l}{\scriptsize Skew = -0.134, Kurtosis = 3.135 (residuals)} &
         & \multicolumn{4}{l}{\scriptsize Skew = -0.129, Kurtosis = 3.234 (residuals)} & \\
\hline
         & \multicolumn{12}{c}{\textbf{Models accounting for publication count ($b$)}} \\
         & \multicolumn{5}{c}{$\alpha = 0.01$ $(N = 1662)$}
         & \multicolumn{5}{c}{$\alpha = 0.05$ $(N = 1685)$}
         & \multicolumn{5}{c}{$\alpha = 0.10$ $(N = 1752)$} \\
\hline
         & Value & SE & $z$ & $p$ & & Value & SE & $z$ & $p$ & & Value & SE & $z$ & $p$ \\
\hline
$c$      & -3.563 & 0.079 & -44.918 & < 0.001 & & -2.591 & 0.143 & -18.175 & < 0.001 & & -2.129 & 0.172 & -12.369 & < 0.001 & \\
$\gamma$ & 0.481 & 0.017 & 28.633 & < 0.001 & & 0.353 & 0.026 & 13.730 & < 0.001 & & 0.290 & 0.029 & 9.971 & < 0.001 & \\
$b$      & -0.0003 & 0.011 & -0.026 & 0.979 & & 0.0014 & 0.009 & -0.155 & 0.876 & & 0.0005 & 0.009 & 0.062 & 0.951 & \\
\hline
\end{tabular}
\begin{tablenotes}
    \centering
    \item $^1$Standard errors robust to cluster correlation (within taxa) were used.
\end{tablenotes}
\end{table}

\newpage
\FloatBarrier
\section*{Network datasets}

All network datasets were downloaded from the Netzschleuder
repository~\cite{peixotoNetzschleuderNetworkCatalogue2020}.
In all analyses only largest connected components were used and networks were
simplified by removing multilinks and self-loops. Moreover, in the case of
directed or weighted networks edge directions and weights were ignored.

Individual network datasets are described below and their unique names within
the repository are provided in the following subsection headings. Each dataset
can be accessed via a generic link of the form:
\texttt{networks.skewed.de/net/<name>}
where \texttt{<name>} is a placeholder which should be substituted with the
name of a specific dataset.

Below we list datasets in groups corresponding to the three empirical analyses
presented in the Main Text. Each section ends with a table with most important
descriptive statistics for all networks. The statistics were calculated for
the largest connected component.

\subsection*{%
    Networks used in
    \enquote{Structural coefficients in real networks}
}

\subsubsection*{%
    Within-organization Facebook friendships
    (\texttt{facebook\_organizations})
}

Six undirected and unweighted networks of friendships among users on Facebook
who indicated employment at one of the target corporations
(S1, S2, M1, M2, L1, L2)~\cite{fireOrganizationMiningUsing2016}.
Companies range in size from small to large.
Only edges between employees at the same company are included in a given
snapshot.

\subsubsection*{Maier Facebook friends (\texttt{facebook\_friends})}

A small anonymized Facebook ego network (undirected and unweighted),
from April 2014~\cite{maierCoverTimeRandom2017}.
Nodes are Facebook profiles, and an edge exists if the two profiles are
\enquote{friends} on Facebook.

\subsubsection*{Facebook ego-network \texttt{(ego\_social)}}

The network is a combination of $10$ ego-nets
sampled from Facebook~\cite{mcauleyLearningDiscoverSocial2012}.
The network is undirected and unweighted.

\subsubsection*{Zachary Karate Club (\texttt{karate})}

Undirected and unweighted network of friendships among members of a university
karate club~\cite{zacharyInformationFlowModel1977}.
We used the corrected version with 78 instead of 77 edges.

\subsubsection*{ANU Residence Hall network (\texttt{residence\_hall})}

A network of friendships among students living in a residence hall at
Australian National University~\cite{freemanExploringSocialStructure1998}.
The original network is directed and weighted with edges indicating that
resident $i$ named resident $j$ as a friend, and weight indicating the level
of friendship: 5 (best friend), 4 (close friend), 3 (friend), 2, 1.

\subsubsection*{Windsurfers network (\texttt{windsurfers})}

A network of interpersonal contacts among windsurfers in southern California
during the Fall of 1986~\cite{freemanHumanSocialIntelligence1988}.
The original network is weighted with weights indicating the perception of
social affiliations majored by the tasks in which each individual was asked
to sort cards with other surfer's name in the order of closeness.

\subsubsection*{Dutch literacy criticism (\texttt{dutch\_criticism})}

A network of criticisms among Dutch literary authors in
1976~\cite{denooyLiteraryPlaygroundLiterary1999}.
The directed edge $(i,j)$ denotes that an author $i$ passed judgment on author
$j$'s work in an interview or review. The original network is also signed and have
positive and negative edges (representing positive and negative judgements).
The edge signs were ignored in our analyses.

\subsubsection*{Advogato trust network (\texttt{advogato})}

A network of trust relationships among users on Advogato, an online community
of open source software developers~\cite{massaBowlingAloneTrust2009}.
Edge directions indicate that node $i$ trusts node $j$,
and edge weight denotes one of four increasing levels of declared trust from
$i$ to $j$: observer (0.4), apprentice (0.6), journeyer (0.8), and master (1.0).

\subsubsection*{Epinions trust network (\texttt{epinions\_trust})}

A who-trusts-whom social network of the general consumer review site
Epinions.com~\cite{richardsonTrustManagementSemantic2003}.
Members can decide whether to \enquote{trust} each other.
These trust relationships are combined with review ratings to determine which
reviews are shown to the user.

\subsubsection*{PGP web of trust (\texttt{pgp\_strong})}

Strongly connected component of the Pretty-Good-Privacy (PGP) web of trust
among users, circa November 2009~\cite{richtersTrustTransitivitySocial2011}.

\subsubsection*{Physician trust network (\texttt{physician\_trust})}

A network of trust relationships among physicians in four midwestern (USA)
cities in 1966~\cite{colemanDiffusionInnovationPhysicians1957}. 
Edge directions indicate that node $i$ trusts or asks for advice from node $j$.
Each of the four components represent the network within a given city.
We analyzed four (disconnected) components corresponding to different cities
as separate networks.

\subsubsection*{E. coli transcription network (\texttt{ecoli\_transcription})}

Directed network of operons and their pairwise interactions, via transcription
factor-based regulation, within the bacteria
Escherichia coli~\cite{shen-orrNetworkMotifsTranscriptional2002}.
In our analyses we used v1.1 version and did not distinguish between different
regulation types.

\subsubsection*{Yeast transcription network (\texttt{yeast\_transcription})}

Directed network of operons and their pairwise interactions, via transcription
factor-based regulation, within the yeast
Saccharomyces cerevisiae~\cite{miloNetworkMotifsSimple2002}.
We did not distinguish between different regulation types.

\subsubsection*{Collins yeast interactome (\texttt{collins\_yeast})}

Undirected and unweighted network of protein-protein interactions in
Saccharomyces cerevisiae (budding yeast), measured by co-complex associations
identified by high-throughput affinity purification and mass spectrometry
(AP/MS)~\cite{collinsComprehensiveAtlasPhysical2007}.

\subsubsection*{Coulomb yeast interactome (\texttt{interactome\_yeast})}

An undirected and unweighted network of protein-protein binding interactions
among yeast proteins~\cite{coulombGeneEssentialityTopology2005}.
Nodes represent proteins found in yeast (Saccharomyces cerevisiae) and an edge
represents a binding interaction between two proteins.

\subsubsection*{Figeys human interactome (\texttt{interactome\_figeys})}

A directed unweighted network of human proteins and their binding
interactions~\cite{ewingLargescaleMappingHuman2007}.
Nodes represent proteins and an edge represents an interaction between two
proteins, as inferred using a mass spectrometry-based
approach.

\subsubsection*{PDZ-domain interactome (\texttt{interactome\_pdz})}

An undirected and unweighted network of PDZ-domain-mediated protein-protein
binding interactions, extracted from the
PDZBase database~\cite{beumingPDZBaseProteinproteinInteraction2005}.
Nodes represent proteins and an edge represents a binding interaction between
two proteins.

\subsubsection*{Joshi-Tope human protein interactome (\texttt{reactome})}

An undirected and unweighted network of human proteins and their binding
interactions, extracted from
Reactome project~\cite{joshi-topeReactomeKnowledgebaseBiological2004}.
Nodes represent proteins and an edge represents a binding interaction between
two proteins.

\subsubsection*{Stelzl human interactome (\texttt{interactome\_stelzl})}

A directed unweighted network of human proteins and their binding
interactions~\cite{stelzlHumanProteinProteinInteraction2005}.
Nodes represent proteins and an edge represents an interaction between
two proteins, as inferred via high-throughput Y2H experiments using
bait and prey methodology.

\subsubsection*{Vidal human interactome (\texttt{interactome\_vidal})}

An undirected and unweighted network of human proteins and their binding
interactions~\cite{rualProteomescaleMapHuman2005}.
Nodes represent proteins and an edge represents a binding interaction between
two proteins, as tested using a high-throughput yeast two-hybrid (Y2H) system.

\begin{table}[ht!]
\centering
\caption{Descriptive statistics ($N = 28$)}
\label{app:tab:stats-domains}
\sffamily
\footnotesize
\begin{tabular}{lllrrlrrrrr}
\toprule
        &             &         &  $s$ &  $c$ &      $n$ &  $S$ &  $\rho$ &  $\langle{d_i}\rangle$ &  $\sigma_{d_i}$ &  $d_{\text{max}}$ \\
domain & dataset & network &      &      &          &      &         &                        &                 &                   \\
\midrule
biological & collins\_yeast &         & 0.62 & 0.01 &     1004 & 0.62 &    0.02 &                  16.57 &            1.12 &            127 \\
        & ecoli\_transcription & v1.1 & 0.02 & 0.05 &      328 & 0.78 &    0.01 &                   2.78 &            1.86 &             72 \\
        & interactome\_figeys &         & 0.01 & 0.14 &     2217 & 0.99 &    0.00 &                   5.79 &            2.95 &            314 \\
        & interactome\_pdz &         & 0.00 & 0.21 &      161 & 0.76 &    0.02 &                   2.60 &            1.12 &             21 \\
        & interactome\_stelzl &         & 0.01 & 0.12 &     1615 & 0.95 &    0.00 &                   3.85 &            1.85 &             95 \\
        & interactome\_vidal &         & 0.04 & 0.03 &     2783 & 0.89 &    0.00 &                   4.32 &            1.63 &            129 \\
        & interactome\_yeast &         & 0.05 & 0.01 &     1458 & 0.78 &    0.00 &                   2.67 &            1.29 &             56 \\
        & reactome &         & 0.61 & 0.05 &     5973 & 0.94 &    0.01 &                  48.81 &            1.39 &            855 \\
        & yeast\_transcription &         & 0.02 & 0.19 &      664 & 0.72 &    0.00 &                   3.21 &            1.79 &             71 \\
social & advogato &         & 0.09 & 0.02 &     5042 & 0.77 &    0.00 &                  15.56 &            2.07 &            803 \\
        & dutch\_criticism &         & 0.16 & 0.22 &       35 & 1.00 &    0.13 &                   4.57 &            0.65 &             12 \\
        & ego\_social & facebook\_combined & 0.52 & 0.02 &     4039 & 1.00 &    0.01 &                  43.69 &            1.20 &           1045 \\
        & epinions\_trust &         & 0.07 & 0.02 &    75877 & 1.00 &    0.00 &                  10.69 &            4.02 &           3044 \\
        & facebook\_friends &         & 0.51 & 0.02 &      329 & 0.91 &    0.04 &                  11.88 &            0.92 &             63 \\
        & facebook\_organizations & L1 & 0.26 & 0.04 &     5793 & 1.00 &    0.00 &                  10.62 &            1.73 &            320 \\
        &             & L2 & 0.22 & 0.02 &     5524 & 1.00 &    0.01 &                  34.11 &            0.93 &            417 \\
        &             & M1 & 0.26 & 0.03 &     1429 & 1.00 &    0.02 &                  27.09 &            1.06 &            339 \\
        &             & M2 & 0.23 & 0.02 &     3862 & 1.00 &    0.01 &                  45.22 &            0.65 &            328 \\
        &             & S1 & 0.29 & 0.03 &      320 & 1.00 &    0.05 &                  14.81 &            0.96 &            113 \\
        &             & S2 & 0.33 & 0.03 &      165 & 1.00 &    0.05 &                   8.80 &            0.95 &             63 \\
        & karate & 78 & 0.26 & 0.06 &       34 & 1.00 &    0.14 &                   4.59 &            0.83 &             17 \\
        & pgp\_strong &         & 0.25 & 0.01 &    39796 & 1.00 &    0.00 &                   9.91 &            2.46 &           1696 \\
        & physician\_trust & 1 & 0.17 & 0.05 &      117 & 1.00 &    0.07 &                   7.95 &            0.50 &             26 \\
        &             & 2 & 0.28 & 0.06 &       48 & 1.00 &    0.16 &                   7.46 &            0.62 &             28 \\
        &             & 3 & 0.32 & 0.08 &       41 & 1.00 &    0.17 &                   6.93 &            0.42 &             15 \\
        &             & 4 & 0.42 & 0.07 &       35 & 1.00 &    0.23 &                   7.83 &            0.48 &             15 \\
        & residence\_hall &         & 0.30 & 0.04 &      217 & 1.00 &    0.08 &                  16.95 &            0.46 &             56 \\
        & windsurfers &         & 0.56 & 0.03 &       43 & 1.00 &    0.37 &                  15.63 &            0.42 &             31 \\
\bottomrule
\end{tabular}
\begin{tablenotes}
    \item $s$ - global similarity (clustering) \\
    \item $c$ - global complementarity \\
    \item $n$ - number of nodes in the giant component \\
    \item $S$ - relative size of the giant component \\
    \item $\rho$ - edge density \\
    \item $\langle{d_i}\rangle$ - average node degree \\
    \item $\sigma_{d_i}$ - coefficient of variation of node degrees \\
    \item $d_{\text{max}}$ - maximum node degree \\
\end{tablenotes}
\end{table}

\FloatBarrier
\subsection*{%
    Networks used in
    \enquote{Similarity and complementarity in social relations}
}

\subsubsection*{Ugandan village networks \texttt{(ugandan\_village)}}

The dataset consists of unweighted and undirected networks of friendship and
health advice relations between households in 17 rural villages bordering
Lake Victoria in Mayuge District, Uganda. It has been originally studied in
Ref.~\cite{chamiSocialNetworkFragmentation2017}. Relations were measured
using the name generator approach in which a representative of each household
was asked to indicate up to 10 persons considered friends or trustworthy in
regard to health issues. Resulting ties were symmetrized.

\begin{table}[ht!]
\centering
\caption{Descriptive statistics ($N = 34$)}
\label{app:tab:stats-social}
\sffamily
\footnotesize
\begin{tabular}{lllrrlrrrrr}
\toprule
        &                 &         &  $s$ &  $c$ &     $n$ &  $S$ &  $\rho$ &  $\langle{d_i}\rangle$ &  $\sigma_{d_i}$ &  $d_{\text{max}}$ \\
domain & dataset & network &      &      &         &      &         &                        &                 &                   \\
\midrule
social & ugandan\_village & friendship-1 & 0.06 & 0.03 &     202 & 1.00 &    0.03 &                   5.42 &            0.72 &             32 \\
        &                 & friendship-2 & 0.11 & 0.05 &     181 & 0.99 &    0.04 &                   7.60 &            0.77 &             44 \\
        &                 & friendship-3 & 0.13 & 0.06 &     192 & 1.00 &    0.06 &                  11.04 &            0.74 &             53 \\
        &                 & friendship-4 & 0.09 & 0.05 &     320 & 1.00 &    0.04 &                  12.97 &            0.66 &             50 \\
        &                 & friendship-5 & 0.12 & 0.05 &     184 & 1.00 &    0.04 &                   7.83 &            0.69 &             30 \\
        &                 & friendship-6 & 0.14 & 0.07 &     139 & 1.00 &    0.07 &                   9.09 &            0.68 &             42 \\
        &                 & friendship-7 & 0.17 & 0.09 &     121 & 1.00 &    0.11 &                  12.73 &            0.52 &             32 \\
        &                 & friendship-8 & 0.06 & 0.04 &     369 & 1.00 &    0.03 &                   9.50 &            0.70 &             58 \\
        &                 & friendship-9 & 0.16 & 0.06 &     178 & 1.00 &    0.07 &                  12.02 &            0.78 &             80 \\
        &                 & friendship-10 & 0.10 & 0.07 &     207 & 1.00 &    0.05 &                  10.55 &            0.64 &             44 \\
        &                 & friendship-11 & 0.09 & 0.04 &     250 & 1.00 &    0.03 &                   8.50 &            0.70 &             44 \\
        &                 & friendship-12 & 0.08 & 0.05 &     229 & 1.00 &    0.03 &                   7.62 &            0.86 &             58 \\
        &                 & friendship-13 & 0.10 & 0.06 &     183 & 1.00 &    0.05 &                   8.84 &            0.68 &             34 \\
        &                 & friendship-14 & 0.15 & 0.06 &     124 & 1.00 &    0.07 &                   8.47 &            0.64 &             36 \\
        &                 & friendship-15 & 0.07 & 0.04 &     120 & 1.00 &    0.04 &                   4.57 &            0.70 &             17 \\
        &                 & friendship-16 & 0.05 & 0.03 &     372 & 1.00 &    0.02 &                   7.38 &            0.72 &             43 \\
        &                 & friendship-17 & 0.25 & 0.07 &      65 & 1.00 &    0.12 &                   7.91 &            0.70 &             31 \\
        &                 & health-advice\_1 & 0.05 & 0.04 &     187 & 0.98 &    0.02 &                   4.61 &            1.31 &             60 \\
        &                 & health-advice\_2 & 0.06 & 0.06 &     170 & 1.00 &    0.03 &                   4.64 &            1.66 &             70 \\
        &                 & health-advice\_3 & 0.08 & 0.05 &     185 & 1.00 &    0.04 &                   6.90 &            1.43 &             78 \\
        &                 & health-advice\_4 & 0.06 & 0.04 &     316 & 1.00 &    0.02 &                   6.61 &            0.96 &             72 \\
        &                 & health-advice\_5 & 0.09 & 0.02 &     166 & 0.99 &    0.03 &                   4.65 &            1.45 &             70 \\
        &                 & health-advice\_6 & 0.06 & 0.02 &     131 & 0.98 &    0.03 &                   4.34 &            1.93 &             75 \\
        &                 & health-advice\_7 & 0.14 & 0.06 &     121 & 1.00 &    0.07 &                   7.82 &            0.80 &             46 \\
        &                 & health-advice\_8 & 0.05 & 0.03 &     361 & 1.00 &    0.02 &                   6.85 &            1.04 &             84 \\
        &                 & health-advice\_9 & 0.10 & 0.05 &     173 & 1.00 &    0.04 &                   7.20 &            1.42 &             98 \\
        &                 & health-advice\_10 & 0.13 & 0.05 &     204 & 1.00 &    0.04 &                   8.04 &            0.98 &             71 \\
        &                 & health-advice\_11 & 0.05 & 0.04 &     234 & 0.97 &    0.02 &                   4.50 &            1.67 &             85 \\
        &                 & health-advice\_12 & 0.04 & 0.04 &     218 & 0.99 &    0.02 &                   4.90 &            1.30 &             79 \\
        &                 & health-advice\_13 & 0.06 & 0.03 &     157 & 0.91 &    0.02 &                   3.64 &            1.25 &             51 \\
        &                 & health-advice\_14 & 0.11 & 0.05 &     120 & 1.00 &    0.05 &                   5.43 &            0.80 &             26 \\
        &                 & health-advice\_15 & 0.06 & 0.05 &     117 & 1.00 &    0.03 &                   3.35 &            1.36 &             34 \\
        &                 & health-advice\_16 & 0.04 & 0.01 &     349 & 1.00 &    0.01 &                   4.94 &            2.21 &            152 \\
        &                 & health-advice\_17 & 0.13 & 0.06 &      63 & 1.00 &    0.07 &                   4.63 &            0.95 &             28 \\
\midrule
    \textbf{} & \textbf{} & \textbf{Average} & \textbf{0.09} & \textbf{0.05} & \textbf{197.29} & \textbf{0.99} & \textbf{0.04} & \textbf{7.21} & \textbf{1.01} & \textbf{56.09}  \\
\bottomrule
\end{tabular}
\begin{tablenotes}
    \item $s$ - global similarity (clustering) \\
    \item $c$ - global complementarity \\
    \item $n$ - number of nodes in the giant component \\
    \item $S$ - relative size of the giant component \\
    \item $\rho$ - edge density \\
    \item $\langle{d_i}\rangle$ - average node degree \\
    \item $\sigma_{d_i}$ - coefficient of variation of node degrees \\
    \item $d_{\text{max}}$ - maximum node degree \\
\end{tablenotes}
\end{table}

\FloatBarrier
\subsection*{%
    Networks used in
    \enquote{Structural diversity across the tree of life}
}

For this analysis we used a dataset of 1840 interactomes of different species
across the tree of life published originally in
Ref.~\cite{zitnikEvolutionResilienceProtein2019}.
The interactomes represent only physical protein-protein interactions
that are experimentally supported or manually curated. Detailed description
of the dataset and its underlying methodology, including the list of types
of interactions that were considered, can be found in the Supplementary
Information appendix of Ref.~\cite{zitnikEvolutionResilienceProtein2019}.
In particular, information on phylogenetic taxonomy information, the source
of publication counts per species and evolution time estimates
defined in terms of the number of nucleotide substitutions per site
are discussed in sections S1.2, S2.2 and S3.

Due to the large number of networks we do not present a table with descriptive
statistics here. The data and code used for conducting the analysis is
available from the Github repository listed in the Main Text
(Data and materials availability).

\section*{
    Experimental assessment of the performance
    of \texttt{PathCensus} algorithm
}

\begin{figure}[htb!]
\centering
\includegraphics[width=.6\textwidth]{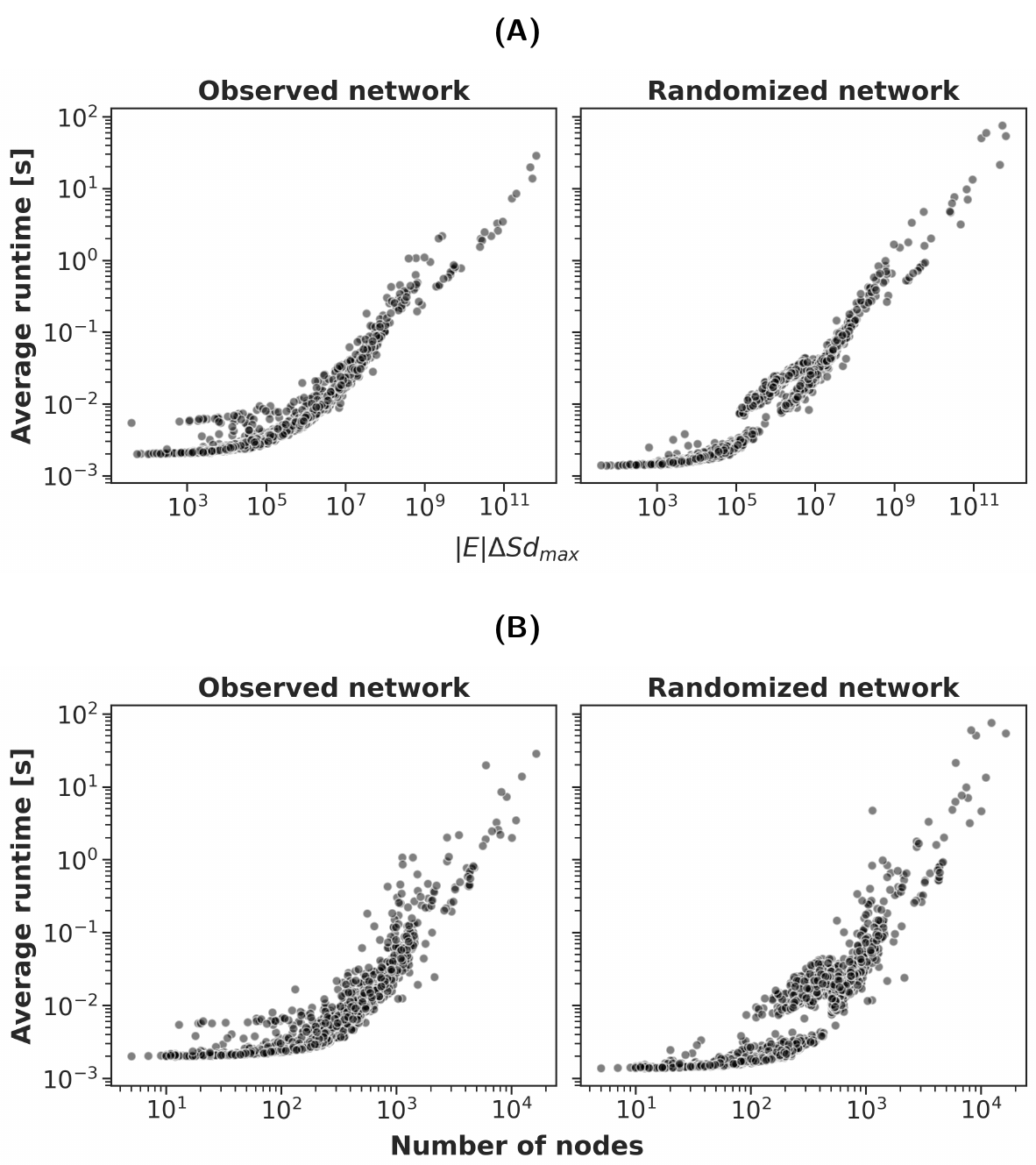}
\caption{
    Experimental assessment of the performance of \texttt{PathCensus}
    algorithm. Computations used the $d_i \leq d_j$ optimization discussed
    in Algorithm~\ref{app:algo:pathcensus}.
    \textbf{(A)}~Average runtime plotted against the product of the number
    of edges ($|E|$), maximum node degree ($d_{\text{max}}$) and the maximum
    size of a $\text{Star}_{i}$ set
    ($\Delta{}S$; see Algorithm~\ref{app:algo:pathcensus}).
    \textbf{(B)}~Average runtime and network size.
}
\label{app:fig:performance}
\end{figure}

We assessed the performance of our implementation of \texttt{PathCensus}
algorithm by measuring the average runtime for each of the 1840 interactome
networks studied in the paper as well as their randomized counterparts
sampled from UBCM. For each network an average
runtime over 5 runs was calculated for the observed network and a corresponding
randomized version sampled from UBCM
(see Fig.~\ref{app:fig:performance}).

Our analysis shows that the runtime scales approximately linearly with
respect to $|E|\Delta{}Sd_{\text{max}}$, which agrees with the previous
theoretical analysis of the computational complexity of \texttt{PathCensus}
algorithm. However, the proportionality constant is lower for smaller
networks and then increases for networks with about 100 nodes.
After that, the linear scaling seems to be stable.

The experiment was run on a machine with Ubuntu (20.04.4 LTS) and
Intel(R) Core(TM) i5-8300H CPU @ 2.30GHz. We did not use the parallelized
version of \texttt{PathCensus} algorithm.

\end{subappendices}

\end{document}